\documentclass[aps,twocolumn,prl]{revtex4}
\newcommand{\be}{\begin{equation}}
\newcommand{\ee}{\end{equation}}

\usepackage{graphicx}
\usepackage{amsmath}
\usepackage{amstext}
\usepackage{amssymb}
\usepackage{amsfonts}
\usepackage{amsbsy} 

\renewcommand{\v}[1]{\boldsymbol{#1}}
\newcommand{\si}{\sigma}
\newcommand{\<}{\langle}
\renewcommand{\>}{\rangle}
\renewcommand{\t}[1]{\tilde{#1}}
\renewcommand{\th}{\theta}
\newcommand{\cH}{{\cal H}}

\begin{document}
\bibliographystyle{apsrev}


\title{Quasi-adiabatic Continuation of Quantum States: 
The Stability of Topological Ground State Degeneracy
and Emergent Gauge Invariance
}
\author{M. B. Hastings$^1$ }
\email{hastings@lanl.gov} 
\author{Xiao-Gang Wen$^2$}
\homepage{http://dao.mit.edu/~wen} 
\affiliation{$^1$Center for Nonlinear Studies and Theoretical Division,
Los Alamos National
Laboratory, Los Alamos, NM 87545}
\affiliation{
$^2$Department of Physics, Massachusetts Institute of Technology,
Cambridge, Massachusetts 02139
}

\date{March 7, 2005}
\begin{abstract}
We define for quantum many-body systems a quasi-adiabatic continuation of
quantum states.  The continuation is valid
when the Hamiltonian has a gap, or else has a sufficiently small
low-energy density of states, and thus is away from a quantum phase
transition.  This continuation takes local operators into local operators,
while approximately
preserving the ground state expectation values.  We apply this
continuation to the problem of gauge theories coupled to matter, and propose a
new distinction, perimeter law versus ``zero law" to identify confinement.
We also apply the continuation to local bosonic models with emergent gauge
theories. We show that local gauge invariance is topological and cannot
be broken by any local perturbations in the bosonic models in either
continuous or discrete gauge groups.  We show that
the ground state degeneracy in emergent discrete gauge theories
is a robust property of the bosonic model, and we argue that the
robustness of local gauge invariance in the continuous case protects
the gapless gauge boson.
\vskip2mm
\end{abstract}
\maketitle

\section{Introduction}

Traditionally,
gauge theory was described as a theory of a vector field $a_\mu$ that has
a local gauge symmetry.  In a Lagrangian framework,
this means $L(a_\mu+\partial_\mu
\phi)=L(a_\mu)$.  This gauge symmetry was believed to be
the defining property
of a gauge theory. It was believed that the gauge symmetry protects the gapless
gauge boson for continuous gauge groups and the topological ground state
degeneracy on
compact space for discrete gauge groups. Even the slightest gauge symmetry
breaking, such as a gauge potential term $(a_\mu)^2$ in the Lagrangian, gives a
finite mass to the gauge boson or lifts the topological degeneracy.  In this
case we can no longer regard the theory as a gauge theory at low energies.

However, the above standard picture for gauge symmetry protecting gapless
gauge bosons and ground state degeneracy is very formal, since the gauge
symmetry is not really a symmetry.
Within the Hamiltonian formulation of gauge theories, the gauge transformation
is simply a transformation between different labels that label the \emph{same}
physical state. It is a do-nothing transformation.  It is very different from
the usual symmetry transformation that transforms a physical state to a
\emph{different} physical state.  
Therefore, it is not clear what is the essence of 
gauge theory and gauge symmetry.

In last 15 years, it was shown, with increasing rigor, that \emph{deconfined}
gauge theories can emerge from certain local bosonic lattice models.
\cite{KL8795,WWZcsp,RS9173,Wsrvb,SF0050,MS0181,Wqoslpub,Wlight,SP0258,BFG0212,%
MS0204,Walight,MS0312,HFB0404,AFF0493,LWstrnet} If the emergent gauge theory
has a continuous gauge group, the local bosonic model will have gapless
excitations that behave just like gauge
bosons.\cite{Wlight,MS0204,Walight,MS0312,HFB0404}  If the emergent gauge
theory has a discrete gauge group or has a Chern-Simons term, the local
bosonic model will have degenerate ground states on compact
space.\cite{Wrig,WNtop,Wsrvb}  This raises a physical question: what protects
the gapless gauge bosons and the ground state degeneracy?  According to the
standard picture for gauge theory, those properties are protected by gauge
symmetry, but from the point of view of local bosonic model, what is this
``gauge symmetry''?  How does ``gauge symmetry'' emerge at low energies?

A close examination of those local bosonic models with emergent gauge theory
reveals that the emergence of gauge theory is intimately connected to
string condensation.\cite{Walight,LWstrnet} The ``gauge symmetry'' is
related to the integrity of the strings.  If the strings are unbreakable, one
can show that the low energy states are gauge invariant.  However, in general,
the strings in the boson model are not perfectly well defined. Strings may
break up momentarily and rejoin.  One may wonder if this means that the
``gauge symmetry'' becomes approximate. We know that a theory that loses its
``gauge symmetry'' even slightly no longer behaves like a gauge theory at low
energies. This seems to suggest that breakable strings will give gauge bosons a
mass gap or lift the ground state degeneracy.

On the other hand, it was believed that the gaplessness of the gauge bosons
and the degeneracy of the ground states in those bosonic models are
topological and are robust against any local
perturbations.\cite{Wrig,WNtop,Wsrvb}
A formal argument goes as the follows (see, for example, Ref. \cite{Wen04},
page 393 and page 435).  We first derive the low energy effective gauge theory
of the bosonic model.  We then argue that any generic perturbation of the
bosonic model cannot generate any terms that break the ``gauge symmetry'' in
the low energy effective gauge theory. Therefore, all the properties protected
by the ``gauge symmetry'' are robust against arbitrary perturbations of the
bosonic model.  

We see that to understand why ``gauge symmetry'' remains exact
even for a generic boson model with virtually breakable strings is vital in
our understanding why the degeneracy of the ground states is protected even
when the original boson model has no symmetry, and why  the gaplessness of the
gauge bosons is protected even when the original boson model has only
translation symmetry.  We would like to address some of these issues in this
paper.

In the next section, we start by introducing some bosonic models with emergent
gauge theories, and raise the question of the ability of topological order and
gauge symmetry to survive local perturbations of the Hamiltonian, as well as
proposing the zero law to identify deconfined gauge theories.  We then
introduce in the following section a quasi-adiabatic continuation which
enables us to identify appropriately dressed operators for the perturbed
Hamiltonian, such that the dressed operator has almost the same ground state
expectation value for the perturbed Hamiltonian as the original operator had
for the original Hamiltonian.  The final section illustrates the application
of the continuation
via a series of examples, beginning with local models without
emergent gauge structure, such as quantum Ising models, and then going on to
emergent gauge theories.  Many of these systems are theories for which the
unperturbed Hamiltonian has a ground state degeneracy and then a gap to the
rest of the spectrum, with no local operators connecting the degenerate ground
states.  We are then able to show that, so long as the gap to the rest of the
spectrum remains open, the splitting between the low energy states of the
perturbed Hamiltonian remains exponentially small.  This is illustrated in the
case of the ferromagnetic quantum Ising model; in the case of the fractional
quantum Hall effect where we are able to extend results on the insensitivity
of the topological degeneracy to disorder\cite{Wrig,WNtop}; and in the case of
an emergent $Z_2$ gauge theory.  The case of gapless theories is also
discussed; it will turn out that gauge symmetry is much more robust
in compact than in non-compact theories.

\section{Simple local bosonic models with emergent gauge theories}

To make the above discussion more concrete, in this section we are going to
discuss two simple bosonic models that have emergent $Z_2$ and $U(1)$ 
gauge theories respectively.

\subsection{A bosonic model with emergent $Z_2$ gauge theory}

The first bosonic model is a spin-1/2 model on a $d$-dimensional cubic lattice.
\cite{W7159,K7959,K032}
The spins live on the links labeled by $\v i$.
The Hamiltonian is given by
\begin{eqnarray}
\label{HZ2}
 H_0=U\sum_{\v I}\left( 1-W_{\v I}\right)
-g\sum_{\v p} \left(\prod_\text{edges of $\v p$} \si^x_{\v i}\right)
\end{eqnarray}
where $\v I$ labels the vertices and $\v p$ the squares of the lattice, and
$W_{\v I}$ is given by
\begin{equation}
\label{xf}
 W_{\v I}=\prod_\text{legs of $\v I$} \si^z_{\v i}.
\end{equation}
The legs of a vertex are the links that connect to the vertex and the edges
of a square are the four links around the square.
$\si^{x,y,z}$ are the Pauli matrices.

\begin{figure}[tb]
\centerline{
\includegraphics[scale=0.4]{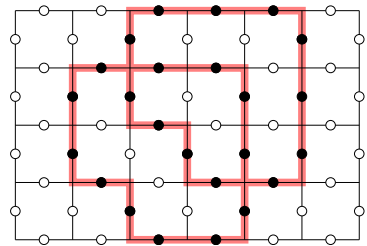}
}
\caption{
A closed-string state. The up-spins are represented by open dots
and the down-spin by filled dots.
}
\label{z2str}
\end{figure}

\begin{figure}[tb]
\centerline{
\includegraphics[scale=0.7]{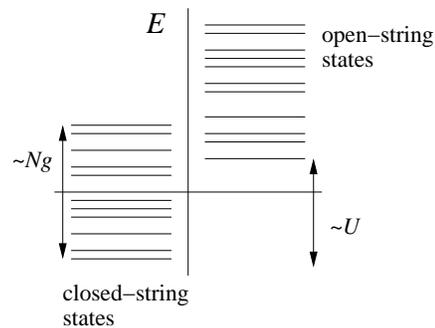}
}
\caption{
The energy levels of $N$ spin-1/2 spins described by $H_0$.
}
\label{specH0}
\end{figure}

When $U$ is very large, the low energy sector of the model is formed by closed
string states. What are the closed string states?
First, the state with all spins up is defined as the no string state.
A closed string state is a state where the down spins form closed loops
(see Fig. \ref{z2str}). 
We note that the close-string states are the states that satisfy
$ W_{\v I} |\text{closed strings}\> = |\text{closed strings}\>$,
Since $[W_{\v I}, H_0]=0$, the closed-string states and the open-string 
states\footnote{The open-string states are states which are not closed-string
states.} do not mix.
This allows us to plot the
spectrum of $H_0$ separately for closed-string states and open-string states
in Fig. \ref{specH0}.

When
$g>0$, the ground state $|\Phi_0\>$ of the model is the equal weight
superposition of all closed string states. Such a state is called a
closed-string condensed state since the closed-string creation operator
$S(C_\text{closed})$ has a non-zero expectation value \cite{Wqoexct}
\begin{eqnarray}
\label{zlaw}
 \<\Phi_0|S(C_\text{closed})|\Phi_0\>=1
\end{eqnarray}
regardless of the size of the string $C_\text{closed}$.
Here a string creation operator that creates a string $C$ is given by
\begin{eqnarray*}
S(C)=\prod_{\v i\text{ on }C}\si^x_{\v i}.
\end{eqnarray*}
We can show (\ref{zlaw}) to be true by noting that
\begin{equation}
\label{SHcom}
 [S(C_\text{closed}), H_0]=0.
\end{equation}

What is the physical character of the closed-string condensed state?  It is
believed that the  closed-string condensed state contain a new kind of order
-- topological order -- which cannot be characterized by symmetry breaking and
long range order.\cite{Wrig,LWstrnet} So we need a new way to characterize such
an order.  One way to characterize the topological order is through the robust
ground state degeneracy on torus.\cite{WNtop,Wsrvb} The closed-string condensed
state in our spin-1/2 model is characterized by a four-fold ground state
degeneracy.

To show the four-fold ground state degeneracy,
we would like to first point out that the $S(C_\text{close})$ are not the only
closed string operators that commute with the Hamiltonian.
We can define dual string operators that also commute with the
Hamiltonian.\cite{K032}

\begin{figure}[tb]
\centerline{
\includegraphics[scale=0.4]{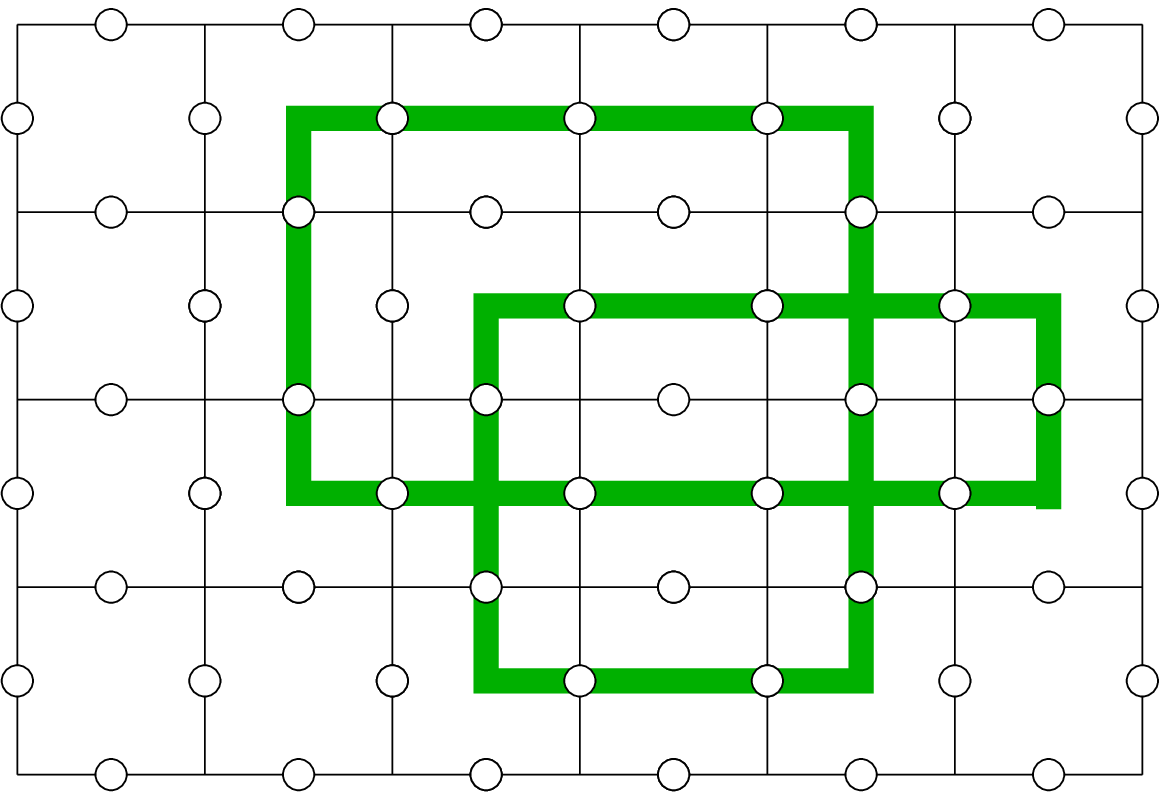}
}
\caption{
A dual closed-string.
}
\label{z2dstr}
\end{figure}

While strings are formed by segments that connect nearest neighbor vertices,
dual strings are formed by segments that connect the centers of
nearest neighbor squares (see Fig. \ref{z2dstr}). 
A dual string operator $\t S(\t C)$ for a dual
string $\t C$ is defined as
\begin{equation}
\label{ds}
\t S(\t C)=\prod_{\v i\text{ cross }\t C}\si^z_{\v i}.
\end{equation}
One can check that
\begin{equation}
\label{SHcomD}
 [\t S(\t C_\text{closed}), H_0]=0,
\end{equation}
which implies that the dual closed-strings also condense
\begin{eqnarray}
\label{zlawD}
 \<\Phi_0|\t S(\t C_\text{closed})|\Phi_0\>=1.
\end{eqnarray}

Now we are ready to show that the ground states of $H_0$ on torus have at
least four-fold degeneracy.  Let $C_x$ and $C_y$ ($\t C_x$ and $\t C_y$)
be the closed (dual) strings that wrap around the torus once in $x$ and $y$
directions.  We find that the four large-string operators
$(S(C_x), S(C_y),\t S(\t C_x), \t S(\t C_y))$ commute with each other
except
\begin{eqnarray}
\label{acr}
 \{S(C_x), \t S(\t C_y)\}=0, \ \ \ \ \ \ 
 \{\t S(\t C_x), S(C_y)\}=0. 
\end{eqnarray}
The above algebra has only one four-dimensional irreducible representation.
Since the large-closed-string operators $(S(C_x), S(C_y),\t S(\t C_x), \t S(\t
C_y))$ act within the degenerate ground states, the ground state degeneracy
must be a multiple of four.

When $U= \infty$, the model becomes the $Z_2$ gauge theory.\cite{W7159}  
One way to see
this is to note that the only states with finite energies are
closed-string states and closed-string states are gauge invariant states under
local $Z_2$ gauge transformations.  The local $Z_2$ gauge transformations are
generated by the unitary operators $W_{\v I}$ in (\ref{xf}):
\begin{eqnarray}
\label{gaugeinv}
 W_{\v I} |\text{closed strings}\> = |\text{closed strings}\> \ \ \
\text{for any }\v I.
\end{eqnarray}
A generic $Z_2$ gauge transformation is given by
\begin{eqnarray}
\prod_{\v I} (W_{\v I})^{n_{\v I}} |\text{closed strings}\> 
= |\text{closed strings}\> .
\end{eqnarray}
In the gauge theory language the closed-string operator
$S(C_\text{closed})$ turns out to be the Wilson-loop operator,\cite{W7159,W7445}
which is gauge invariant
\begin{equation*}
W_{\v I} S(C_\text{closed}) W_{\v I}^\dag = S(C_\text{closed}) .
\end{equation*}

(\ref{zlaw}) implies that the expectation values
of the Wilson loop satisfies the perimeter law
\begin{eqnarray}
\label{plaw}
 \<\Phi_0|S(C_\text{closed})|\Phi_0\>\sim e^{-\alpha |C_\text{closed}|}
\end{eqnarray}
with zero coefficient $\alpha=0$. Here $|C_\text{closed}|$ is the length
of the string $C_\text{close}$.
In this case we will call (\ref{zlaw}) the zero law.

The perimeter law indicates that the $Z_2$ gauge theory is in the deconfined
phase. The $Z_2$ deconfined phase has four nearly degenerate ground states on
the torus which is  consistent with our previous direct calculation on the
spin model. The energy separation between the four nearly degenerate ground
states is of order $e^{-L/\xi}$ where $L$ is the linear size of the torus and
$\xi$ a finite length scale.

Now let us add a term
\begin{equation*}
 H_1=-J_1\sum_{\v i} \si^z_{\v i}
\end{equation*}
to our spin model Hamiltonian $H_0$ and assume $U$ is finite.  
$S(C_x)$ and $S(C_y)$ no longer commute
with the modified Hamiltonian $H_0+H_1$. So it is not clear if $H_0+H_1$ still
has four degenerate ground states on a torus. 


To understand the properties of modified spin system $H_0+H_1$, we note that
$H_0+H_1$ still does not mix the closed-string and open-string states.  So if
$U\gg g,J_1$, the low lying states are still closed string states.  One can
check that if we restrict $H_0+H_1$ to the close-string subspace, the system
is identical to a pure lattice $Z_2$ gauge theory.
A pure $Z_2$ gauge theory has two phases in d+1 dimensions if $d > 1$: a
deconfined phase where the expectation value of the Wilson loop satisfies the
perimeter law, and a confined phase where the expectation value of the Wilson
loop satisfies the area law: \begin{eqnarray} \label{alaw}
\<\Phi_0|S(C_\text{closed})|\Phi_0\>\sim e^{-\gamma A(C_\text{closed})},
\end{eqnarray} where $A(C_\text{closed})$ is the area enclosed by the loop
$C_\text{closed}$ and $\gamma >0$.  So we expect our spin model $H_0+H_1$ also
has two phases.  If $|J_1|\ll |g|$, the ground state of the spin model is
filled with large closed strings which correspond to the $Z_2$ deconfined
phase.  If $J_1 \gg |g|$, the ground state of the spin model hardly has any
strings (i.e. almost all spins point up) which correspond to deconfined phase.

Base on this picture, we guess that when  $|J_1|\ll |g|$, the ground states of
$H_0+H_1$ are (nearly) four-fold degenerate, while when $|J_1|\gg |g|$ the
ground state is not degenerate.  But the result for $|J_1|\ll |g|$ clearly is
just a guess. Can we provide a more rigorous proof?

\begin{figure}[tb]
\centerline{
\includegraphics[scale=0.7]{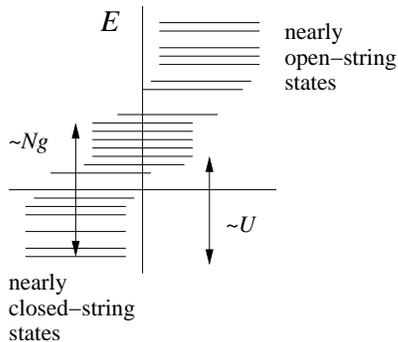}
}
\caption{
The energy levels of $N$ spin-1/2 spins described by $H_0+H_1+H_2$, assuming
$|J_1|,|J_2| \ll |g| \ll U \ll N|g|$.
}
\label{specH012}
\end{figure}

The situation gets even more complicated if we add another term
\begin{equation*}
 H_2=-J_2\sum_{\v i} \si^x_{\v i}.
\end{equation*}
Such a term can mix the closed-string and open-string states.  The low energy
sector of $H_0+H_1+H_2$ is not formed by simple closed-string states, although
the mixing with the open string states may be small for small $J_2$ (see Fig.
\ref{specH012}).  It appears that $H_2$ breaks the $Z_2$ ``gauge symmetry''
since the low energy states no longer satisfy (\ref{gaugeinv}).

Also when $J_2 \neq 0$ and $U<\infty$, the expectation
value of a closed string operators satisfies the perimeter law in both
$|J_1|\ll |g|$ limit and $J_1\gg |g|$ limit, which give no sign of two phases.
All of those suggest that the four-fold ground state degeneracy is lifted
by a finite $J_2$.

\begin{figure}[tb]
\centerline{
\includegraphics[scale=0.7]{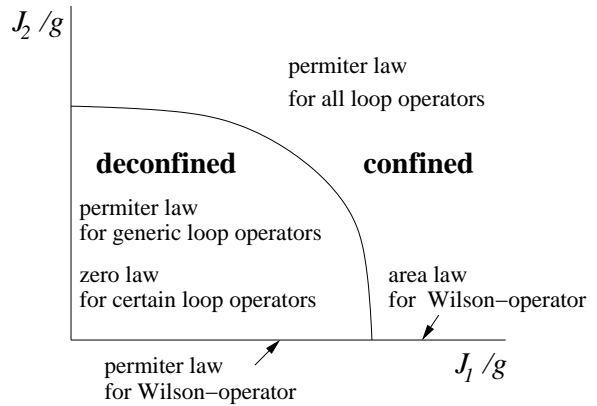}
}
\caption{
A likely quantum phase diagram for the spin-1/2 system $H_0+H_1+H_2$.
The deconfined phase is characterized by
four nearly degenerate ground states on torus
with energy splitting of order $e^{-L/\xi}$ where $L$ is the linear size of the
torus and $\xi$ a length scale.
The confined phase is characterized by a non degenerate ground state on torus.
In general, the confined phase and the deconfined phase are distinguished by
the zero law and the perimeter law of certain loop operators.
}
\label{z2phs}
\end{figure}

But this suggestion is incorrect. Ref.  \cite{Wsrvb} argues that any local
perturbations of $H_0$ cannot break the $Z_2$ ``gauge symmetry''.  As a
result, $H_0+H_1+H_2$ will have four nearly degenerate ground states as long
as $J_1$ and $J_2$ are not too large. Such a phase contains a non-trivial
topological order. However, Ref.  \cite{Wsrvb} only provides a formal
argument. A more rigorous understanding is needed.

Certainly, large $J_1$ and $J_2$ will polarize the spins and lift the ground
state degeneracy.  Such a phase has a trivial topological order.  This
suggests a phase diagram in Fig. \ref{z2phs}.  We note that the two phases in
Fig. \ref{z2phs} have the same symmetry and are distinguished only by
topological orders.  

{}From the phase diagram, we see that when $J_2\neq 0$ the perimeter/area laws
of the closed string operators (or Wilson loop) cannot determine if the ground
state is topologically ordered or not.  Thus for a generic bosonic model,
perimeter/area laws of the closed string operators is not the proper way to
test if the ground state has closed-string condensation (or non-trivial
topological order).  

In the examples
section, we will show that the topological phase is characterized
by dressed closed-string operators which satisfy the zero law. The trivial
phase does not contain any such closed string operator.  The closed string
operators in the trivial phase all satisfy the perimeter law.  So it is the
zero/perimeter laws that distinguish topological/trivial (deconfined/confined)
phases, instead of perimeter/area laws.

For $J_1=J_2=0$, the system has an exact four-fold degeneracy of the ground
state on the torus, followed by a gap of order $g$ to the next lowest state.
The closed-string operators satisfy the zero law.  In the examples section, we
will show that, for small but non-zero $J_1$ and $J_2$, a deformed or dressed
closed-string operator can still  satisfy the zero law.  The zero law of the
dressed string operator allows us to show the four-fold degeneracy of the
ground states in the small $J_{1,2}$ limit.  More precisely, we assume that
the gap, from the four lowest states to the rest of the spectrum, remains
open, and then we show the four-fold ground state degeneracy up to an
exponentially small splitting.  We will also show that, in the small $J_{1,2}$
limit, the low energy sector of the model is still formed by $Z_2$ gauge
invariant states.  However, for $J_2\neq 0$, the gauge invariance is under a
deformed $Z_2$ gauge transformation.  So in this sense, none of the small
perturbations in the spin model can break the ``gauge symmetry'' in the low
energy effective gauge theory.

\subsection{A bosonic model with emergent $U(1)$ gauge theory}

In our second bosonic model, we consider rotors on the links of
$d$-dimensional cubic lattice.
A rotor can be viewed as a particle moving on a circle.
The position of the particle is given by an angle $\th$,
and the angular momentum of the particle by $L^z=-i \partial_\th$.
The Hamiltonian of the rotor model is given by
\begin{align}
\label{strnetH}
H_\text{rotor} &= U\sum_{\v I}  Q^2_{\v I}
- g \sum_{\v p} (B_{\v p}+h.c.)
\nonumber\\
&\ \ \ \ + J_1 \sum_{\v i} (L^z_{\v i})^2 
+ J_2 \sum_{\v i} (L^+_{\v i}+L^-_{\v i})
\\
B_{\v p} &= L^+_{1}L^-_{2}L^+_{3}L^-_{4},\ \ \ \ \ \ \ \ \ 
Q_{\v I} = (-)^{\v I} \sum_{\text{legs of }\v I} L^z_{\v i} ,
\nonumber 
\end{align}
where $\v I$ 
labels the vertices, $\v i$ labels the links and $\v p$
labels the squares of the cubic lattice.
1, 2, 3, 4 label the four links that form
the edges of the square $\v p$.
$L^+=e^{i\th}$ is the raising operator of $L^z$, $L^-=(L^+)^\dag$, and
$(-)^{\v I}=1$ for the even vertices and.
$(-)^{\v I}=-1$ for the odd vertices.


\begin{figure}[tb]
\centerline{
\includegraphics[scale=0.7]{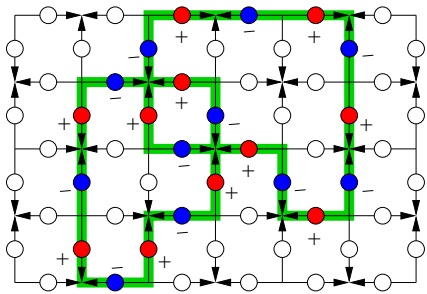}
}
\caption{
The empty dots represent rotors with $L^z=0$ -- the no string states. 
A closed string is formed by a loop connecting
neighboring vertices.
A closed-string state is obtained by
alternately increase or decrease $L^z$
by 1 along the closed string.
The filled dots represent rotors with $L^z=\pm 1$.
The arrows on the links all point from even vertices to odd vertices.
}
\label{u1str}
\end{figure}

When $g=J_1=J_2=0$ and $U>0$, 
the ground states are highly degenerate and form a low
energy subspace.  One of the ground states is the state with $L^z_{\v i}=0$
for every rotor.  Other ground states can be constructed from the first ground
state by drawing an loop in the cubic lattice, and then alternately increasing
or decreasing $L^z$ by 1 along the loop (see Fig.~\ref{u1str}).  The sum
$\sum_{\text{legs of }\v I} L^z_{\v i}$ vanishes on every vertices for such a
closed-string state.  Such a process can be repeated to construct all of the
degenerate ground states.  We see that the degenerate ground states are formed
by loops, or more precisely string-nets, since loops can overlap and form
branched strings.


The $J_1$-term gives the strings in the degenerate ground states a fine
energy and represents string tension. The $B_{\v p}$ operator creates a small
loop of closed string or deforms the existing strings. 
Thus the $g$-term generates string ``hopping'' or string fluctuations.

The $J_{1,2}$-term and the $g$-term lift the degeneracy of the ground states.
In the $U\gg J_{1} \gg \text{max}(|g|,|J_2|)$ limit, the true ground state correspond to a
state with almost no strings (i.e., a state with $L^z_{\v i}=0$ on every
link). The excitations above such a state have finite energy gaps.  In the
$U\gg |g| \gg J_{1,2}>0$ limit, the true ground state is a superposition of
many large closed strings \cite{Walight,Wen04}. Such a state is a string-net
condensed state.  

When $J_2=0$, the Hamiltonian $H_\text{rotor}$ does not mix the closed-string
states and open-string states. 
When  restricted to the closed-string
subspace, $H_\text{rotor}$ is identical to the Hamiltonian of lattice $U(1)$
gauge theory.
The closed-string states can be viewed as gauge invariant states
since they are invariant under local $U(1)$ gauge transformations
\begin{equation}
\label{u1gauge}
 e^{i \phi Q_{\v I}}
|\text{closed-string}\> = |\text{closed-string}\>,\ \ \ \
\text{for any } \v I.
\end{equation}
A general $U(1)$ gauge transformation is generated by
$e^{i\sum_{\v I} \phi_{\v I} Q_{\v I}}$.

In the limit $|J_1| \gg |g|$, the lattice $U(1)$ gauge theory is in the strong
coupling limit and is in a confined phase.  In the limit $|J_1| \ll |g|$, the
lattice $U(1)$ gauge theory is in the weak coupling limit and has gapless
$U(1)$ gauge bosons as its only low lying excitations if $d>2$.  So when
$J_2=0$ and when $U\gg |g| \gg |J_1|$, the rotor model $H_\text{rotor}$
contains emergent gapless $U(1)$ gauge
bosons.\cite{MS0204,Walight,MS0312,HFB0404,Wen04}

When $J_2\neq 0$, $H_\text{rotor}$ mixes the closed-string and open-string
states. The low energy states are no longer pure closed-string states and is
not invariant under the local $U(1)$ gauge transformation (\ref{u1gauge}). It
appears that a non-zero $J_2$ will break the $U(1)$ gauge symmetry. We may
conclude that even a small $J_2$ will give the $U(1)$ gauge boson a gap and
the rotor model $H_\text{rotor}$ cease to have emergent $U(1)$ gauge bosons at
low energies.  In the section of examples,
we will show that this line of argument
is incorrect.  For a small $J_2$ (or any other small perturbations to
$H_\text{rotor}$), we can define deformed local $U(1)$ gauge transformations
so that the low energy states of $H_\text{rotor}$ are invariant under the
deformed local $U(1)$ gauge transformations.  Thus the local ``$U(1)$ gauge
symmetry'' cannot be broken by any small perturbations if $U\gg |J_{1,2}|,
|g|$.  Thus far, we will prove these results for $U(1)$ theories;
we do not prove, but strongly conjecture, that the stability of
the $U(1)$ gauge symmetry protects the gaplessness of the
gauge boson.
As a result, no translation invariant perturbation can give the gapless
$U(1)$ gauge boson a mass gap. The gaplessness of the emergent $U(1)$ gauge
boson is topologically stable!

\section{Quasi-Adiabatic Continuation}

In this section we define the quasi-adiabatic continuation.  We consider a
family of Hamiltonians, ${\cal H}_s$, depending on a continuous parameter $s$,
where we wish to define a continuation from $s=0$ to $s=1$.  We denote
eigenstates of ${\cal H}_{s}$ by $\Psi_{a,s}$, where a state $\Psi_{a,0}$
evolves into state $\Psi_{a,s}$ under an {\it adiabatic} evolution of ${\cal
H}_s$.  In the event of a level crossing as a function of $s$, any
arbitrary continuation of the states through the level crossing is allowed.

Let us begin with some motivation and discussion.  We define the unitary
operator $V(s)=\sum_a |\Psi_{a,s}\rangle \langle\Psi_{a,0}|$.  Then, for any
operator $O$, we could define a corresponding dressed operator by $O_{\rm
adiab}(s)=V(s)OV(s)^{\dagger}$ so that $O_{\rm adiab}(s)$ would have exactly
the same expectation value in state $\Psi_{a,s}$ as $O$ does in state
$\Psi_{a,0}$.  Using such a definition of dressed operators, we can show that
the dressed string operators and the dressed gauge transformations will have
the same properties in the deformed model as the bare string operators and the
bare gauge transformations in the exactly soluble model. However, this
definition of a dressed operator would not suffice for our purposes at all!
In particular, we do not have any reason to believe that the resulting $O_{\rm
adiab}(s)$ would still be a local operator.  

Indeed, even if $O$ only involves operators on a few sites, in general, the
continued $O_{\rm adiab}(s)$ will contain operators on every site of the
system.  However, in this section, we will show that, under certain
conditions, we can continue $O$ into an local operator that acts only on a
finite number of sites.  

To state the result more precisely, let us first assume that $\cH_s$ has a gap
$\Delta E$
separating a low energy sector and a high energy sector for all $0<s<1$.  
For
any operator $O$ which acts on a set of sites $S_O$, we will construct
an approximate dressed operator $O(s)$ which only acts on sites within
a distance $l$ from the sites $S_O$.  Then, there is a unitary matrix 
$Q_0(s)$ which acts only within the low energy sector of $\cH_s$, such that
\begin{equation}
\label{boundGap}
 \<\Psi_\text{low,s}|Q_0(s)^{\dagger} O(s) Q_0(s)
- O_{\rm adiab}(s)
|\Psi_\text{low,s}\> < N_{S_{O,l}} e^{- l/\xi}
\end{equation}
in the large $l$ limit for a certain
fixed length scale $\xi$, where $N_{S_{O,l}}$ is the number of
sites within distance $l$ of a site in $S_O$.  This number grows
only as a power of $l$, so is easily overwhelmed by the exponential.
The state $|\Psi_\text{low,s}\>$ is any state in the low energy sector
of $\cH_s$.  

If $\cH_s$
has no gap between the low energy and high energy
sectors, we will define a density of states $\rho(E)$ in the high
energy sector.  If $\rho(E)$ is
bounded by $\rho(E)< E^{\alpha-1}$, then the 
bound in (\ref{boundGap}) is weakened to
\begin{equation}
\label{boundNoGap}
 \<\Psi_\text{low,s}|Q_0(s)^{\dagger} O(s) Q_0(s)-O_{\rm adiab}(s) 
|\Psi_\text{low,s}\> < 
N_{S_{O,l}} l^{1-\alpha/2}.
\end{equation}
For a local operator $O$, $N_{S_{O,l}}$ grows as $l^d$, where
$d$ is the dimension of the lattice, and so if $d+1-\alpha/2<0$ then
the error decays for large $l$.  For a string-like operator,
$N_{S_{O,l}}$ is proportional to $l^{d-1} |O|$, where $|O|$ is the
length of the strength, for $l<|O|$, and is proportional to $l^d$ for
$l>|O|$.
(A more rigorous statement of the results will be given later.)

The key in obtaining the above result is to
adopt a {\it different} definition of the dressed operator by
$O(s)=\t V(s)O \t V(s)^{\dagger}$, where the unitary operator $\t V(s)$ is defined
following Eq.~(\ref{weq}).  Physically, the definition of $V$ corresponds to
adiabatically changing the Hamiltonian from ${\cal H}_0$ to ${\cal H}_s$,
while the definition (\ref{weq}) of $\t V(s)$ corresponds to a quasi-adiabatic
change of the Hamiltonian.  So $\t V(s)$ can be viewed as an approximation of
$V(s)$.  The operator $\t V(s)$ will be chosen to achieve the goal of defining
dressed operators $O(s)$ which have approximately the same ground state
expectation values in the perturbed Hamiltonian ${\cal H}_s$ as the original
operators did in the unperturbed Hamiltonian, while preserving the locality of
the operators.

The locality of the operators is very important.  Suppose, for example that
the ground state of ${\cal H}_0$ has long-range correlations.  That is, there
exist two local operators, $O_1,O_2$ which are correlated even though the
two operators are far separated from each other in space.  For example, if
this is a spin system with long-range spin correlations these
may be spin operators acting on two different sites which are far separated from
each other.  Then, however, the operators $O_1(s),O_2(s)$ will also
be correlated in the ground state of ${\cal H}_s$, and since the operators
remain local under the continuation this implies the existence of long-range
correlations in the ground state of ${\cal H}_s$.

The major result will be an explicit definition of $\t V(s)$ in terms of
derivatives of the Hamiltonian which accomplishes these goals.  In order to
show that ground state expectation values remain approximately unchanged under
the continuation, we will make some assumptions on the existence of a gap,
though some extensions to gapless systems with sufficiently small low-energy
density of states will be discussed.  The dressed operators remain local under
this evolution: local operators are spread out over a length scale of order
the correlation length in a gapped system, while string-like operators are
spread out over a length scale which grows only logarithmically with the
string length.  Detailed proofs will be given in an Appendix.

The importance of the low energy density of states in what follows can be
understood physically by analogy to another continuation that should be much
more familiar, namely Fermi liquid theory.  As discussed by
Anderson\cite{pwa}, the correct way to think of Fermi liquid theory is to
think of starting with a non-interacting system and turning the interactions
on slowly, but not infinitely slowly; that is, physically exactly the same
procedure we imagine here.  Anderson's discussion of how fast the interactions
need to be turned on is based on considerations of the quasi-particle states,
and corresponds very closely to our two criteria: maintaining both the
expectation values and the locality of the operators.  To show that this is
possible, the analysis in the Fermi liquid case relies on the low density of
particle-hole excitations near the Fermi surface.  Here, we rely on something
similar, namely a low density of states at low energy.  Our continuation is
very general, and thus valid for a much wider range of systems than Fermi
systems, but this generality can in some cases limit what we can prove on
specific systems.

\subsection{Definition of System}

We consider a family of Hamiltonians, ${\cal H}_s$, which obey the
finite-range conditions\cite{LR7251,H0402}: ${\cal H}_s=\sum_i {\cal H}_s^i$,
where letters $i,j,...$ labels different lattice sites of the system; each
${\cal H}_s^i$ acts only on sites $j$ with $d(i,j)\leq R$ where $R$ is the
interaction
range and $d(i,j)$ is some metric on the lattice; and $||{\cal H}_s^i||\leq J$
for some constant $J$ for all $i,s$.  We further assume that
$||\partial_s{\cal H}_s^i|| \leq K$ for some constant $K$ for
all $i,s$.  It is possible to slightly
weaken the finite-range conditions and consider exponentially decaying
interactions as well\cite{LR7251}.

To define the concept of ``density of states'', let us focus on a state
$\Psi_{\rm low,s}$
that we continue.  We will consider a $\Psi_{\rm low,s}$ which is a
ground state of $\cH_s$, but in general we could continue any eigenstate of
$\cH_s$.  The state $\Psi_{a,s}$ has energy $E_a>0$ compared to $\Psi_{\rm
low,s}$ if $a$ is not in the low energy sector.
We assume that, for $0\leq s \leq 1$, the
density of states of the local operator $u^i_s=\partial_s {\cal H}^i_s$ is
bounded as follows.  We assume that there is a $D(E)$ such that $\sum_{a\in
high,
|E_a| \leq E}|\langle \Psi_{a,s}|u^i_s| \Psi_{\rm low,s}\rangle|^2\leq D(E)
||u^i_s||$, where $||...||$ denotes the operator norm.  The sum is restricted
to states $\Psi_a$ in the high energy sector.
Note that $D(E)\leq 1$
for all $E$.  We define $\partial_E D(E)=\rho(E)$ to be the density of states
at energy $E$ produced by operator $u^i_s$.

There are a number of systems for which this density of states bound is
relevant.  For a discrete gauge theory, there will be a set of topological
excitations below a gap.  These topological excitations
form the low energy sector, and thus $D(E)$ vanishes below the gap.
For a transverse-field Ising system, ${\cal H}^i=J
\sum_{<i,j>} \si^z_i \si^z_j+B \si^y_i$, in the ferromagnetic phase, there
will again be two almost degenerate states below a finite gap.  At $B=0$, these
states correspond to symmetric and anti-symmetric combinations of all spins up
or all spins down.

The bound $D(E)$ implies a locality bound\cite{H0402}.  Suppose some operator
$O$ acts only on some set of sites, $S_O$.  We define the distance between a
site $j$ and the operator $O$ to be equal to $d(j,O)={\rm min}_{i\in
S_O}(d(i,j))$, with the minimum is taken over sites $i \in S_O$, and we define
the distance between two operators $O_1,O_2$ to be $d(O_1,O_2)={\rm min}_{i\in
S_{O_1},j \in S_{O_2}}(d(i,j))$.  Suppose the system has a unique ground
state.  Then if the system is gapped, so that $D(E)=0$ for $E\leq \Delta E$,
the connected expectation value $|\langle O u^i_s \rangle_s -\langle O
\rangle_s \langle u^i_s \rangle_s|$ is exponentially decaying in $l$, where
$\langle ... \rangle_s$ denotes the ground state expectation value with
Hamiltonian $\cH_s$.  On the other hand, if $D(E)\propto E^{\alpha}$, then
$|\langle O u^i_s \rangle_s -\langle O \rangle_s \langle u^i_s \rangle_s|$ is
bounded by some constant times $d(i,O)^{-\alpha/2}$.

\subsection{Definition of Quasi-Adiabatic Continuation} 

We introduce the unitary operator
\begin{eqnarray}
\label{weq}
&&\t V(s) \\
&=&{\cal S}'\exp\left\{-\int_0^{s}
{\rm d}{s'}
\int_0^{\infty} {\rm d}\tau 
e^{-(\tau/t_q)^2/2}
[\tilde u_{s'}^+(i\tau)-
h.c.]
\right\}, \nonumber 
\end{eqnarray}
where the symbol ${\cal S}'$ denotes that the exponential is $S'$-ordered, in
analogy to the usual time ordered or path ordered exponentials.  We define
$u_{s}=\partial_s{\cal H}_{s}=\sum_i u^i_s$, and define $\tilde
u_{s}^+(i\tau)$ following Ref. \cite{H0402}: for any operator $A$
\begin{eqnarray} 
{\tilde A}(t)\equiv A(t) 
\exp[-(t/t_q)^2/2]
, \\
\label{tdef}
{\tilde A}^{\pm}(\pm i\tau)
=\frac{1}{2\pi}
\int {\rm d}t \,
{\tilde A}(t)\frac{1}{\pm it+\tau}.
\end{eqnarray}
The time $t_q$ will be chosen later.  For comparison with previous
work\cite{H0431,H0402}, $t_q$
is what was previously called $\sqrt{q}/\Delta E$; in this work, we also
consider the possibility of gapless theories where there is no scale $\Delta
E$.  The time evolution of operators is defined by $A(t)=\exp[i{\cal H}_{s'}
t]A\exp[-i{\cal H}_{s'}t]$.  The Hermitian conjugate in Eq.~(\ref{weq}) of
$\tilde u_{s}^+(i\tau)$ is $\tilde u_{s}^-(-i\tau)$, and $\t V(s)$ is a unitary
operator.

At $t_q=\infty$, the operator $\t V(s)$ becomes equal to
$V(s)=\sum_a\Psi_{a,s}\rangle \langle\Psi_{\rm low,s}$.  To see this, note that at
$t_q=\infty$, we have $\tilde A^+(i\tau)=A^+(i\tau)$, where $A^+(i\tau)$ is
the positive energy part of $A$ taken at imaginary time $i\tau$.  That is, in
a basis of eigenstates of ${\cal H}_s$ with energies $E_a,E_b$, we have matrix
elements $A^+(i\tau)_{ab}= A_{ab}\Theta(E_a-E_b)\exp[-\tau(E_a-E_b)]$, where
$\Theta(x)$ is the step function.  Then, $-\int_0^{\infty} {\rm d}\tau
[u^+(i\tau)-u^-(-i\tau)]= -(E_a-E_b)^{-1} \partial_s{\cal H}_s$, which gives
the result of linear perturbation theory for the adiabatic evolution of
quantum states with a change in the Hamiltonian.  We instead keep $t_q$ finite
to define a ``quasi-adiabatic" evolution, which will transform local operators
into local operators.  To show that keeping $t_q$ finite maintains the
locality we will rely on finite group velocity results, while we will use the
gap to show that we get only a small error in the ground state expectation
values by taking a finite $t_q$; detailed proofs of this are in the Appendix,
while the physical discussion is given in the next section.
We will relate the time $t_q$ to the scale $l$ by Eq.~(\ref{lscale}) below.

\subsection{Results}
\label{res}

For any operator $O$, we define $O(s)=\t V(s) O \t V(s)^{\dagger}$, where
$O(s)$ has been ``smeared out" over a scale $l$ given by Eq.~(\ref{lscale})
below.  For a gapped theory, we only need to take a length $l$ of order the
correlation length $\xi$ to get a small error in Eq.~(\ref{boundGap}).  To
understand how this works, define $Q(s)=\tilde V(s) V(s)^{\dagger}$.  Then,
for a state $| \Psi_{\rm low,s} \rangle$ in a low energy subspace, $\langle
\Psi_{\rm low,s} | O(s) | \Psi_{\rm low,s} \rangle= \langle \Psi_{\rm low,0} |
Q(s) O Q^{\dagger}(s) | \Psi_{\rm low,0} \rangle$.  
If we can show that $Q(s)$ has nearly vanishing matrix elements
between the low energy and high energy states,
we can bound the difference in expectation values in Eq.~(\ref{boundGap}).
The gap between the states in the low energy subspace and the high energy
subspace enables us to show this; for a given gap $\Delta E$, if we perform
the quasi-adiabatic continuation sufficiently slowly, we can show that
(loosely speaking) the matrix elements of $Q(s)$ between the low energy and
high energy states almost vanishes and $Q(s)$ almost acts within the low
energy subspace as an unitary operator.

However, there is one complication: in a
macroscopic system, $Q(s)$ produces excitations throughout the sample.  Thus,
for infinitesimal $s$, $Q(s)=1+s \sum_i e^i + ...$, where the
anti-Hermitian operator $e^i$ creates
local excitations near site $i$.  The probability that $Q(s)$ produces some
excitation acting on the ground state diverges with the system size.  
Strictly speaking, the matrix elements of $Q(s)$ between the low energy and
high energy states are not small.
However,
the terms $e^i$ in $Q$ with $i$ sufficiently far from $O$ can be commuted
through $O$ and do not affect the final expectation value.
We will show that
the error from the terms $e^i$ with $i$ near $O$ is
exponentially small in $l$ divided by the correlation length.
The proof in the appendix follows this argument, using a triangle inequality
to make the bound precise.  

If $D(E)$ is not gapped, but has a power law behavior, $D(E)\propto
E^{\alpha}$ for small $\alpha$, then in a $d$-dimensional system we can still
find a local $O(s)$ if $\alpha/2>d+1$ according to
Eq.~(\ref{boundNoGap}).  This requirement on the exponent can
be physically understood as follows: if we keep $t_q$ finite so that $O(s)$ is
spread out over a length scale $l$ under the quasi-adiabatic evolution, then
we have to worry about any error in $Q(s)$ on the length scale $l$.
The correlation
function of $u_s$ with $O$ decays as $l^{-\alpha/2}$, and this
also is how the error terms decay, while the space-time
volume at scale $l$ is of order $l^{d+1}$.  Thus, this
requirement can be understood in terms of the relevance or irrelevance of
$u_s$ at large scales in the given correlation function. 

This physical description is based on two technical results.  First, we
claim that $O(s)$ is {\it local} up to a scale $l$.  Specifically, we claim
that for any operator $O_j$ which acts only on site $j$, 
\begin{eqnarray}
\label{le}
||[O_j,O(s)]||\leq X_1 ||O_j|| ||O|| \times \\ \nonumber
{\rm max}(\exp[-d(j,O)/\xi_C],\exp[-(d(j,O)/l_q)^2/2]),
\end{eqnarray}
for some constant $X_1$ of order unity and constant $l_q$ of order $t_q/c_1$,
where $c_1$ is a characteristic inverse velocity of the system.  The length
$\xi_C$ is a microscopic length of order the interaction range of the
Hamiltonian and is defined later.  Eq.~(\ref{le}) implies that, while
$O(s)$ does involve sites more than a distance $l_q$ from $S_O$,
the commutator becomes exponentially small once $d(j,O)$ becomes larger than
$l_q$.  To relate $l_q$ and $l$, we ask for the commutator to
be smaller than the error in Eq.~(\ref{boundNoGap}) or Eq.~(\ref{boundGap})
once $d(j,O)$ becomes larger than $l$.
For a gapless theory, we only need to take
$d(j,O)$ logarithmically larger than $l_q$ before the exponential decay becomes
much smaller than $l^{d+1-\alpha/2}$; in this case, $l$ is of order $l_q$.
For a gapped theory, we need that $\exp[-(l/l_q)^2]\sim \exp[-l/\xi]$, so
that 
\be
\label{lscale}
l\sim l_q^2/\xi \sim t_q^2 \Delta E/c_1.
\ee
The correlation length $\xi$ is at most of order $1/(c_1 \Delta E)$.

If one prefers to have an operator which involves only sites
within a distance $l$ of $S_O$,
and {\it exactly} commutes with sites more
than a distance $l$ from $S_O$, one may
define
an operator $O_{trunc}(s)$ such that
\begin{widetext}
\begin{eqnarray}
\label{otb}
||O_{trunc}(s)-O(s)||\leq X_2 
\sum_{j, d(i,j)\geq l}
{\rm max}(\exp[-d(j,O)/\xi_C],\exp[-(d(j,O)/l_q)^2/2]),
\end{eqnarray}
\end{widetext}
for some constant $X_2$ of order unity.  In order for the error
in Eq.~(\ref{otb}) to be of order the error in Eq.~(\ref{boundNoGap}) or
Eq.~(\ref{boundGap}), we need to pick $l$ as above: $l\sim l_q^2/\xi$ in
the gapped case, up to logarithmic corrections.

Second, we claim that $O(s)$ has almost the same expectation value as
$O_{\rm adiab}(s)$, up to the unitary matrix $Q_0$.
Specifically, we show that for $0\leq s \leq 1$
\begin{eqnarray}
\label{ce}
|\langle \Psi_{\rm low,s}|
Q_0(s)^{\dagger} O(s) Q_0(s) -O_{\rm adiab}|\Psi_{\rm low,s}\rangle|
\\ \nonumber
\leq
2 ||O|| [c_2(s)+c_3(s)],
\end{eqnarray}
where $c_2(s),c_3(s)$ are given by Eqs.~(\ref{teq},\ref{difb}) in
the Appendix.
One may verify from the calculation in the Appendix that
the error term, $c_2(s)+c_3(s)$, gives the error described above in
Eqs~(\ref{boundGap},\ref{boundNoGap}).
Note that if
$O=O_1 O_2 ... O_n$, then $O(s)=O_1(s) O_2(s) ... O_n(s)$, so the expectation
values of products of operators are also approximately preserved under this
quasi-adiabatic evolution.
Thus,
$|\langle \Psi_{\rm low,s}|Q_0(s)^{\dagger} O_1(s) O_2(s) ... Q_0(s)-
O_{1,{\rm adiab}}
O_{2,{\rm adiab}} ...  |\Psi_{\rm low,s} \rangle| \leq
2 ||O_1|| ||O_2|| ... [c_2(s)+c_3(s)]$, where here the $c_2(s),c_3(s)$
are the error terms appropriate for the product operator $O_1O_2...$

\section{Examples}
We illustrate the quasi-adiabatic continuation by a series of
examples.  We start with local operators, considering a system with
$Z_2$ (Ising) symmetry, a fractional quantum Hall system,
and then a system with $U(1)$ symmetry.  We then
repeat the process in the case of non-local, string operators.

\subsection{Local Ising Model}
The first and simplest example is a quantum Ising
ferromagnet
in a transverse field.  Let the Hamiltonian be
${\cal H}=J \sum_{<i,j>} \si^z_i \si^z_j+B \sum_i \si^y_i$
where each site has a spin-$1/2$ and
the ferromagnetic interaction $J$ couples nearest neighbor spins on the
lattice.

For $B=0$, the system has two exact ground states, one state which
we denote $\Psi_{\uparrow}$ with all spins
up and one state $\Psi_{\downarrow}$ with all spins down.
The gap to the lowest excited state above these two states
is $2Jq$ where $q$ is the coordination
number of the lattice; this state has one flipped spin.  
We define symmetric and anti-symmetric combinations
$\Psi_{S,A}=(1/\sqrt{2})(\Psi_{\uparrow}\pm \Psi_{\downarrow})$.  The states
$\Psi_{S,A}$
are eigenvectors of the operator
$\prod_i \si^y_i$, with eigenvalues $\pm 1$.  This operator,
which flips the spin on every site, commutes with the Hamiltonian for
all $B$.  

We now consider the quasi-adiabatic continuation with
${\cal H}_s=
J \sum_{<i,j>} \si^z_i \si^z_j+s B \sum_i \si^y_i$.  At $s=0$, this
has the exact ground states $\Psi_{S,A}$.
For $sB$ small enough, the two lowest eigenstates of ${\cal H}_s$
are adiabatic continuations of
$\Psi_{S,A}$.  Hence,
the matrix element of the operator $\partial_s {\cal H}_s=
B \sum_i \si^y_i$
between these two states vanishes for all $B$, since
$\partial_s {\cal H}_s$ commutes with $\prod_i \si^y_i$ and these
two states are eigenstates of $\prod_i \si^y_i$ with different eigenvalues.

Above these two states, there is a gap to the rest of the spectrum.  It is
known that for small enough $B$, the gap will remain open for all $s$ with
$0\leq s\leq 1$.  Thus, we can perform the continuation.  The
vanishing of the matrix elements of $\partial_s \cH_s$ between the
two low energy states implies that $Q_0(s)$ is equal to the identity
matrix.  We consider the
continuation of the operator $\si^z_i$.  The ground state for $B=0$ may be
taken to be $\Psi_S$.  For $B=0$, we have $\langle \Psi_S |
\si^z_{i_1} \si^z_{i_2} ...
\si^z_{i_n} | \Psi_S \rangle=1$ for $n$ even and $-1$ for $n$ odd.
The quasi-adiabatic
continuation $\si^z_i \to \si^z_i(s)$ spreads
out the $\si^z_i$ operators over a distance $l$.  Certainly $l$ is less
than the linear size of the system.
Then $\langle \Psi_{S,s} |
\si^z_{i_1}(s) \si^z_{i_2}(s) ... \si^z_{i_n}(s) |\Psi_{S,s}
\rangle$,
again equal to $1$ or $-1$, up to some error, depending on whether $n$ is even
or odd.  Here, $\Psi_{S,s}=V(s)\Psi_S$.
The error is exponentially small in $l/\xi$.  We conjecture that
the operators $\si^z_{i_1}(s)$ correspond to {\it block spin} operators: they
are equal to plus or minus $1$ depending on whether the average spin over a
correlation volume is positive or negative.

If we instead consider the Hamiltonian
${\cal H}_s=J \sum_{<i,j>} \si^z_i \si^z_j+
s B \sum_i \si^z_i$, the operator $\partial_s {\cal H}_s$ will have 
non-vanishing
matrix elements between the states $\Psi_{S,A}$, and thus we must start
with states $\Psi_{\uparrow},\Psi_{\downarrow}$ to perform the continuation
if we want to have $Q_0(s)$ be equal to the identity.
In this case, $\si^z_i(s)=\si^z_i$ and $\langle \Psi_{\uparrow,s} | 
\si^z_{i_1}(s)
\si^z_{i_2}(s) ... \si^z_{i_n}(s) | \Psi_{\uparrow,s} \rangle=(+1)^n$, 
where we have assumed that the sign of $B$ is such that $\Psi_{\uparrow,s}$
is the ground state, rather than $\Psi_{\downarrow,s}$.  For the opposite
sign of $B$, the correlation function is instead $(-1)^n$.

The two cases, depending on the different ways to add the magnetic field
transverse or parallel to the $z$-axis, lead to different ground
state correlation functions.  In the second case, the correlation
function of the continued operator
is not a very interesting result; the unit operator would have
the same correlation function.  However, in the first case, the ability
to find a continuation of the operator with the given correlation functions
is a much more interesting result.  There are long-range correlations
in the operator, since $\langle \Psi_{S,s}|
\si^z_{i_1}(s) \si^z_{i_2}(s) |\Psi_{S,s} \rangle-
\langle \Psi_{S,s}|
\si^z_{i_1}(s) |\Psi_{S,s} \rangle \langle \Psi_{S,s} | 
\si^z_{i_2}(s) |\Psi_{S,s}\rangle$ is
non-vanishing even when sites $i_1,i_2$ are far from each other, and
in particular even when
the distance between $i_1,i_2$ much larger than $l$.  This
implies\cite{H0402} the presence of another state close in energy to the
ground state when the magnetic field is added in the transverse
direction.

However, we would like to show the ground state degeneracy
in another way, based directly on continuing the states
$\Psi_{S,A}$.  This way will be very valuable for
more complicated systems such as the quantum Hall system.
We will consider a more general class of Hamiltonians ${\cal H}_s$:
we consider arbitrary Hamiltonians which are sums of local
terms, and which commute with $\prod_i \si^y_i$.
We show that, under the assumption that the gap between the
two lowest states and
the rest of the spectrum remains open, the energy difference
between the continuation of the two lowest
states is exponentially small.  As long
as that gap remains open, it is possible to adiabatically continue
these two states, giving states $\Psi_{S,s}=V(s)\Psi_{S}$ and
$\Psi_{A,s}=V(s)\Psi_{A}$ as eigenstates
of ${\cal H}_s$.  Here, we rely on the fact that ${\cal H}_s$
commutes with $\prod_i \si^y_i$ and thus has vanishing
matrix elements between the states.

To compute
the difference in energies for given $s$, we compute
$\langle \Psi_{A,s} |{\cal H}_s |\Psi_{A,s} \rangle-
\langle \Psi_{S,s}  |{\cal H}_s |\Psi_{S,s} \rangle=
\langle \Psi_{A} |V(s)^{\dagger} {\cal H}_s V(s) |\Psi_{A} \rangle-
\langle \Psi_{S} |V(s)^{\dagger} {\cal H}_s V(s) |\Psi_{S} \rangle$.
Thus, $V(s)^{\dagger} {\cal H}_s V(s)$ defines
the continuation of the operator $\cH_s$ from $s$ back to $0$ and
$\t V(s)^{\dagger} {\cal H}_s \t V(s)$ defines a quasi-adiabatic
continuation from $s$ back to $0$:
$\langle \Psi_{S,s} | {\cal H}_s | \Psi_{S,s} \rangle$
is approximately equal to
$\langle \Psi_{S} | \t V(s)^{\dagger} {\cal H}_s \t V(s) | \Psi_{S} \rangle$.
The error in this continuation
is exponentially small in $l/\xi$.  The operator
${\cal H}_s$ is a sum of local operators, while the operator
$\t V(s)^{\dagger} {\cal H}_s \t V(s)$ is a sum of terms spread
out over length scale $l$.  The only operators $O$ such that
$\langle \Psi_{A}| O |\Psi_{A} \rangle-
\langle \Psi_{S}| O |\Psi_{S} \rangle\neq 0$ are operators which
flip every spin in the system and thus have non-vanishing
matrix elements between $\Psi_{\uparrow}$ and $\Psi_{\downarrow}$.
Thus all {\it local} operators have vanishing matrix elements
between the two states $\Psi_{\uparrow}$ and $\Psi_{\downarrow}$.  
In particular, if the length scale
$l$ is smaller than the system size $L$, then
$\langle \Psi_{A}| \t V(s)^{\dagger} {\cal H}_s \t V(s) |\Psi_{A} \rangle=
\langle \Psi_{S}| \t V(s)^{\dagger} {\cal H}_s \t V(s) |\Psi_{S} \rangle$.
Thus we can pick $l$ just smaller then the system size to
show that 
$\langle \Psi_{A}| V(s)^{\dagger} {\cal H}_s V(s) |\Psi_{A} \rangle-
\langle \Psi_{S}| V(s)^{\dagger} {\cal H}_s V(s) |\Psi_{S} \rangle$
is of order $||{\cal H}_s|| \exp(-L/\xi)\sim L^d \exp(-L/\xi)$.
Here, the bound for this system is not a very tight bound: one
expects the level splitting to be exponentially small in
$(L/\xi)^d$ instead.

The key steps in this argument were that $(1)$: matrix
elements of operators which commute with $\prod_i \si^y_i$ vanish
between $\Psi_{S}$ and $\Psi_{A}$; and $(2)$: all
local operators have the same expectation values in state
$\Psi_A$ as in state $\Psi_S$.

\subsection{Quantum Hall Effect}

We now turn to the case of the fractional quantum Hall effect.
Consider a system with no disorder on a torus at filling factor
$p/q$, with $p$ and $q$ coprime.  The magnetic translation
group implies at least a $q$-fold degeneracy of the ground
state\cite{WNtop}.  Assume that in the absence of disorder
there is a $q$-fold degenerate ground state, with a gap
to the rest of the spectrum.  Now, consider adding disorder to the
system, defining ${\cal H}_s={\cal H}_0+s \int dx dy
U(x,y) \Psi^{\dagger}(x,y) \Psi(x,y)$, where $U(x,y)$ is
a disorder potential and ${\cal H}_0$ is the clean Hamiltonian.
Wen and Niu argued\cite{WNtop} that to first order in $s U(x,y)$ the
splitting between the $q$-fold degenerate states was exponentially
small.  We will use the continuation to show that the splitting
is exponentially small for non-vanishing disorder strength
under the sole assumptions that
the gap to the rest of the spectrum remains open and
that at $s=0$ all local operators have the same expectation value
up to exponentially small terms in the $q$ lowest states
and that at $s=0$ all local operators have exponentially small
matrix elements between the $q$ lowest states.  We note
that since these last statements involve only $s=0$ they
can be checked in specific model systems without disorder, and
in fact were checked
in \cite{WNtop} when they showed the exponentially small
splitting of the $q$ lowest states in linear perturbation theory.

Suppose, then that for $0\leq s \leq 1$ there are
$q$ states, $\Psi_{n,s}$ for $n=0...q-1$,
and a gap to the rest of the
spectrum.  We wish to show that these states are close
in energy.  Thus, we compute
$ \langle \Psi_{m,s} | {\cal H}_s |\Psi_{m,s} \rangle-
\langle \Psi_{n,s} | {\cal H}_s  |\Psi_{n,s} \rangle$,
for some $m,n=0...q-1$.
As above, $V(s)^{\dagger} {\cal H}_s V(s)$ defines
the continuation of the operator $\cH_s$ from $s$ back to $0$, and
$\t V(s)^{\dagger} {\cal H}_s \t V(s)$ defines a quasi-adiabatic
continuation from $s$ back to $0$.

However, unlike the case of the
quantum Ising model, we do not have any symmetries to make
matrix elements of ${\cal H}_s$ and $\partial_s {\cal H}_s$ vanish
between the low lying states.  Thus, we do not have any control on
the matrix $Q_0(s)$.
Then,
$\langle \Psi_{m,s} | {\cal H}_s | \Psi_{m,s} \rangle$
is equal to
$\langle \Psi_{m,0} | Q_0(s)^{\dagger} \t V(s)^{\dagger} {\cal H}_s \t V(s) Q_0(s) | \Psi_{m,0} \rangle$,
up to an error of order $||{\cal H}_s|| \exp(-l/\xi)$.  Since we
are continuing from non-zero $s$ to $s=0$, now the matrix $Q_0$ acts
within the low energy sector of $\cH_0$.

However, this matrix $Q_0(s)$ causes no problem.  As before, for $l<L$,
$\t V(s)^{\dagger} {\cal H}_s \t V(s)$ is a local operator, and then
under the assumptions above, the expectation value
$\langle \Psi_{m,0} | Q_0(s)^{\dagger} \t V(s)^{\dagger} {\cal H}_s \t V(s) Q_0(s) | \Psi_{m,0} \rangle$ is independent of $Q_0(s)$, up to
exponentially small corrections, thus giving the desired result.

\subsection{Local Rotor Model}
The next example is a system with a continuous symmetry.  We take the
Hamiltonian ${\cal H}_s= k(s)^{-1}\sum_i \Pi_i^2 + k(s)\sum_{<i,j>}
z_i \overline{z_j}$, where $z_i$ is a continuous complex field with
$|z_i|=1$, and $\Pi_i$ is a
momentum with $[z_i,\Pi_i]=i z_i$.  The parameter $k(s)$ is an $s$-dependent
stiffness of the field $z$.

We pick a quasi-adiabatic continuation using $k(s)=k_0 (k_1/k_0)^s$, so
$k(0)=k_0$ and $k(1)=k_1$, thus $\partial_s {\cal H}_s=
\ln(k_1/k_0)
[-k(s)^{-1} \sum_i \Pi_i^2
+k(s)\sum_{<i,j>} z_i\overline{z_j}]$.  We choose the initial
$k_0$ to be large compared to the final $k_1$.
We assume that we $k_1$ is sufficiently large that the system
is still in a phase with gapless excitations and algebraic
correlations.  In this phase, we can compute the
density of states by writing $z=\exp(i\phi)$ for some $\phi$
with a Gaussian action for $\phi$.
At low energy, the system acquires relativistic
invariance, and thus in $d$ dimensions the density of single particle
states at energy $E$ is of order $E^{d-1}$.  The
matrix element of $\phi$ between the ground state and such a
single particle state is of order $E^{-1/2}$, and thus the
integrated density of states created by $\phi$ is
$D(E)\propto E^{d-1}$.
The integrated density
of states below energy $E$ created by $\partial_s {\cal H}_s$ instead
is $KD(E)\sim \ln(k_0/k_1) E^{2d+2}$.  Then,
$\alpha/2=d+1$ and we are in a marginal case: the error in
the continuation is of order $\log(L/l)$.

\subsection{Emergent Discrete Gauge Theories}

We now turn from these theories with local operators to emergent
gauge theories.  We consider the Hamiltonian of the emergent $Z_2$
gauge theory in the introduction.

For $J_1=J_2=0$, we have 4 exactly degenerate ground states on a torus, and a
gap to the rest of the spectrum.  For non-zero $J_1,J_2$, the deformed
Hamiltonian is still local $\cH_s=\sum_{\v i} \cH_s^{\v i}$, where $\cH_s^{\v
i}$ is a local operator defined near the site $\v i$.  We can use the same
reasoning used in FQH states to show that $\cH_s$ still has 4 exactly
degenerate ground states on a torus.  
However, here we will use a slightly different approach.  Continuing $\cH_s$
from $s$ back to 0, we find that 
\begin{equation*}
\<\Psi_{m,s}|\cH^{\v i}_s|\Psi_{n,s}\> =
\<\Psi_{m,0}|V^\dagger(s)\cH^{\v i}_sV(s)|\Psi_{n,0}\>, 
\end{equation*}
where $|\Psi_{n,s}\>$, $n=1,...,4$, are the 4 low lying states of $\cH_s$.
Due to the energy gap separating the 4 low lying states with rest of states,
the operator $V^\dagger(s)\cH^{\v i}_sV(s)$ is almost a local operator.  More
precisely, up to an error of order $e^{-l/\xi}$ and a unitary rotation $Q_0(s)$
between the 4 low lying states $|\Psi_{n,0}\>$, 
$V^\dagger(s)\cH^{\v i}_sV(s)$ can be replaced
by a truncated operator $H^{\v i}$ which only acts on sites within a distance
$l$ from the site $\v i$.  Thus we have 
\begin{equation*}
\<\Psi_{m,s}|\cH^{\v i}_s|\Psi_{n,s}\> = 
\<\Psi_{m,0}|Q_0(s)^\dagger H^{\v i} Q_0(s)|\Psi_{n,0}\>
+O(e^{-l/\xi}).  
\end{equation*}
Since $Q_0(s)|\Psi_{n,0}\>$ are the ground states of the exactly soluble model
(\ref{HZ2}), they from a irreducible representation of the algebra of the
large-closed-string operators (\ref{acr}).  We can choose the length
$l$ over which the operators are smeared to be one
quarter of the linear size $L$ of the system.  In this case $H^{\v i}$ will be
local enough that we can choose the positions large-closed-string operators to
avoid any overlap between $H^{\v i}$ and the closed-string operators.  So
$H^{\v i}$ commutes with the closed-string operators.  As a result, $H^{\v i}$
must be proportional to the identity operator within the irreducible
representation.  This way, we have shown that $\<\Psi_{m,s}|\cH^{\v
i}_s|\Psi_{n,s}\>\propto \delta_{mn}$ up to an error $e^{-L/4\xi}$, which
implies that $\<\Psi_{m,s}|\cH_s|\Psi_{n,s}\>\propto \delta_{mn}$ up to an
error $L^2e^{-L/4\xi}$.  So the energy splitting between the 4 low lying
states is less than $L^2e^{-L/4\xi}$ for the deformed Hamiltonian $\cH_s$.


We can also continue any closed-string or dual closed-string
operators $S(C_\text{closed})$ and
obtain
\begin{equation*}
\<\Psi_{m,0}|S(C_\text{closed})|\Psi_{n,0}\> =
\<\Psi_{m,s}|V(s)S(C_\text{closed})V^\dagger(s)|\Psi_{n,s}\>, 
\end{equation*}
Then there exist dressed
closed string operators $S^\text{dre}(C_\text{closed})$
that have a width $l$ such that
\begin{align*}
&\<\Psi_{m,s}|Q_0(s)^\dagger S^\text{dre}(C_\text{closed}) Q_0(s)|\Psi_{n,s}\> 
\nonumber\\
=&
\<\Psi_{m,s}|V(s)S(C_\text{closed})V^\dagger(s)|\Psi_{n,s}\>
+O(e^{-l/\xi}) , 
\end{align*}
Here $Q_0(s)$ is a unitary rotation between the 4 low lying states
$|\Psi_{n,s}\>$ and $Q_0(s)$ is independent of the closed-string operators.
This implies that the dressed string and dual string operators
$S^\text{dre}(C_\text{closed})$ have the same algebra among the low energy
states of the perturbed Hamiltonian as the original operators
$S(C_\text{closed})$ do for the original Hamiltonian (up to an error
$\exp(-l/\xi)$).

The above also implies that the expectation of the dressed closed-string
operators in the ground state of the perturbed Hamiltonian satisfies the zero
law (\ref{zlaw}), up to an error of order $|S| \exp(-l/\xi)$, where $|S|$ is
the string length. We see that the error is exponentially small for long
strings since $l$ can be chosen to be a fraction of $|S|$.


Our final task is to show the invariance of certain states under a deformed
gauge transformation for $J_2\neq 0$.  First consider $J_1,J_2=0$.  The
Hamiltonian has the 4 ground states which are invariant under the local gauge
transformation $W_{\v I}$ given by Eq.~(\ref{xf}).  We can create gauge
invariant excited states by acting on the 4 ground states by open dual string
operators.  Following Eq.~(\ref{ds}), we define an open dual string operator
as $\t S(\t C)=\prod_{\v i\text{ cross }\t C}\si^z_{\v i}$, where now the dual
string
$\t C$ is open, with two endpoints.  The states created by acting with $\t S(\t
C)$ are still invariant under the local $Z_2$ gauge transformations, since
$[\t S(\t C),W_{\v I}]=0$.  They are excited states which introduce $Z_2$
gauge flux at the endpoints of the open string.  By acting on the ground state
with the open dual strings, we can create all gauge invariant states, and for
$U\gg g$, these are the lowest energy excited states.  Continuing to non-zero
$J_1$ at $J_2=0$ still leaves these states invariant under the local $Z_2$ gauge
transformation.  Continuing to non-zero $J_2$ breaks the local $Z_2$ gauge
invariance.  However, these states are invariant under the deformed
local $Z_2$ gauge transformation $W_{\v I}(s)$.
To see this, we use the continuation to show that $\langle
\Psi_{\rm low,s}| W_{\v I}(s) |\Psi_{\rm low,s} \rangle=1$, up to exponentially small error in $l/\xi$.  Since
$W_{\v I}(s)$ is unitary, this implies the gauge invariance of the ground
state under the deformed gauge transformation up to exponentially small error.
The gauge invariance, up to the same small error, of the states created by
acting on the ground state with operators $\t S(\t C,s)=\t V(s) \t S(\t C,s)
\t V(s)^{\dagger}$ then follows from the exact commutator $[\t S(\t C,s),W_{\v
I}(s)]=0$.  The size of the deformed gauge group is, however, much smaller
than the original: if the linear system size is $L$, there are $(L/l)^d$
different local gauge transformations rather than $L^d$ as is the case at
$J_2=0$.

\subsection{Emergent Continuous Gauge Theories}

Our final problem is the theory with the $U(1)$ gauge symmetry,
Eq~(\ref{strnetH}).  The absence of a gap makes it much more difficult
to obtain results on this system.  We will obtain only one result,
the existence of a deformed $U(1)$ gauge invariance of the system.

We consider some gauge transformation acting on a site,
$W_{\v I,\phi}=e^{i \phi Q_{\v I}}$.  When $J_2=0$, the ground state
has an exact gauge invariance: $\langle \Psi_{0}| W_{\v I,\phi} |\Psi_0
\rangle=1$.  If
we continue to non-zero $J_2$, the ground state breaks the gauge invariance.
However, we claim that the expectation
value of the continued operator, $\langle \Psi_{0,s}|W_{\v I,\phi}(s) |
\Psi_{0,s} \rangle$,
is still close to unity.
Unlike the $Z_2$ case, this does {\it not} follow simply from the results
derived previously as there is no gap, and the low energy density of states
is too high to use the results for gapless systems.  However, it is still
possible to show this result.  We only very briefly sketch the argument,
leaving a more detailed presentation for future work.
The idea is as follows: at $J_2=0$, the system
has gauge non-invariant states at an energy of order $U$ above the ground state.
Under the continuation, at finite $t_q$, the operator $\t V(s)$ will take
the ground state at $s=0$ into some state which is a mix of ground and
excited states.  That is, the expectation value 
$\langle \Psi_{0,s} | W_{\v I,\phi}(s) | \Psi_{0,s}
\rangle=\langle \Psi_0 | Q(s)^{\dagger} W_{\v I,\phi}
Q(s)| \Psi_0
\rangle$, where the unitary matrix $Q(s)=\tilde V(s)^{\dagger}V(s)$
mixes the ground state with the excited states.  However, since all the
low-lying excited states at $J_2=0$ are gauge-invariant, it is no longer
necessary to show that $Q(s)^{\dagger} W_{\v I,\phi} Q(s)$ is close to
$W_{\v I,\phi}$; 
it suffices to show that we do not mix in the states which
are not gauge invariant under $W_{\v I,\phi}$ and
all such states are at energy of order $U$ above the ground state.  At
infinitesimal $s$, the mixing into such states is exponentially small, since
the energy $U$ acts like a gap; however, one has some mixing into
the low-lying gauge-invariant states.  As $s$ increases, one mixes
into progressively higher energy states, until eventually one
begins to excite the gauge non-invariant states.  However, if $t_q$
is sufficiently big (roughly of order $U^{-1}$) the mixing into the
gauge non-invariant states can be bounded at $s=1$, thus obtaining
the desired result.

Now consider the excited states.  At $J_2=0$,
the gauge invariant
excited states of the system can be obtained by acting on the
ground state with operators of the form $O=\prod_{i} e^{i \phi_{i}
L^z_i}$, where the phase $\phi_{i}$ is some arbitrary function of the
leg.  These operators commute with $W_{\v I,\phi}$ and therefore at
non-zero $J_2$ the
continued operators commute with the continued gauge transformation:
$[O(s),W_{\v I,\phi}(s)]=0$.  Therefore, there are a class of excited
states for non-zero $J_2$, namely the excited states created by
acting on the ground state with the operators $O(s)$, which are
also gauge invariant under the deformed
gauge transformation, up to the same error.

As in the $Z_2$ case, the size of the deformed gauge group is much smaller
than the original gauge group.  The original gauge group had
$L^d$ different generators, while the deformed group has only
$(L/l)^d$ such generators.
A more careful analysis should be able to then use this deformed
gauge invariance to show that the gapless photon is protected.  This is also
a job for the future.

The compactness of the gauge group was important here.  Consider a non-compact
$U(1)$ theory, with Lagrangian $(1/2)(\partial_{\mu}A_{\nu}- \partial_{\nu}
A_{\mu})^2+(\lambda/2) (\partial_\mu A^\mu)^2+M A_\mu A^\mu$.  The term in
$\lambda$ is a gauge fixing term and $M$ is a term which breaks the gauge
symmetry.  With this quadratic action, it is easy to verify that a non-zero
$M$ opens a gap.  Why doesn't the continuation work here?  The reason is that
this theory has no gap to the gauge non-invariant states, unlike the compact
cases before, despite the gauge fixing term.  The gauge fixing term does not
open a gap to the gauge non-invariant states in this case, instead it adds
{\it gapless} longitudinal and scalar photons to the theory.  To say it
differently, in a compact theory, the charge is quantized and the open string
states have a minimum possible energy because they necessarily terminate in an
end with a charge that is a multiple of the charge quantum.  In a non-compact
theory, the charge may be arbitrarily small. So the energy cost is also
arbitrarily small.

\section{Discussion}
The standard wisdom is that as long as gaps remain open, a system
does not have a quantum phase transition and thus the long-distance
structure of correlation functions remains the same.  We have shown
a precise form of this statement.  We have found that, by appropriately
dressing operators, the long-distance structure can in fact be
preserved to a much greater degree than one might have expected.
In particular, we have shown the presence of the zero law (\ref{zlaw})
for gauge theories in the deconfined phase, and we propose this
as a test of confinement.  Further, we have considered the stability
of topological order under perturbations of the Hamiltonian, and shown
that the order is robust unless the gap to the rest of the spectrum
(local excitations) closes.

Topologically ordered states are described by emergent gauge theories at low
energies. The topological order is closely tied to the emergent gauge
invariance of the low energy gauge theories. From the point of view of the low
energy gauge theory, our result shows that the  emergent low energy gauge
invariance is topological. It cannot be broken by any local perturbations in
the parent bosonic model. We hope this result will shed light on the true
meaning of gauge invariance and gauge theory.

The continuation is also useful for systems without emergent local gauge
structure, and may have a wider applicability.  For example, an outstanding
question is to prove that in some neighborhood of the AKLT
point\cite{AKL8799,AKL8877}, a spin-$1$ chain remains gapped\cite{haldane}.
The continuation might be useful in doing this, or at least in showing, under
the assumption of the existence of a gap, the persistence of string
order\cite{string} throughout the Haldane phase.

{\it Acknowledgments---}
MBH was supported by DOE contract W-7405-ENG-36.  
XGW was supported by NSF Grant No. DMR--04--33632,
NSF-MRSEC Grant No. DMR--02--13282, and NFSC no. 10228408.

\appendix
\section{Proof of Locality and Approximation Results}

\subsection{Locality Result}
To show Eq.~(\ref{le}),
we use the finite group velocity result, proven in \cite{LR7251,H0431}.
This result uses
the finite-range conditions on the Hamiltonian above to
bound the commutator
$||[A(t),B(0)]||$, where $A(t)=\exp(i{\cal H}_s t)A\exp(-i{\cal H}_s t)$.  One
can
show that this commutator is exponentially small for times $t$ less than
$c_1 l$ where $l$ is the distance between $A$ and $B$ and $c_1$ is
some characteristic inverse velocity which depends on $J$, $R$, and
the lattice structure.
The specific bound is that
$||[A(t),B(0)|| \leq ||A|| ||B|| \sum_j g(t,d(A,j))$,
where the sum ranges over sites $j$ which
appear in operator $B$ and where the function $g$ has the property that
for $|t|\leq c_1 l$,
$g(c_1 l,l)$ is exponentially decaying in $l$ for large $l$ with
decay length $\xi_C$ for some constant $\xi_C$ which is of order $R$.
Recall that $d(A,j)$ is the minimum over sites $i$ acted on by
$A$ of the distance $d(i,j)$.

Before giving the proof of Eq.~(\ref{le}), we give a physical description.  
The finite group
velocity result has a very simple interpretation.  Consider a local operator
$A$.  Under time evolution, we get an operator $A(t)$ which ``spreads out"
over space as time passes.  The finite group velocity result implies that
for finite $t$ the operator $A$ is still local up to some
length $t/c_1$ (here $c_1^{-1}$ is some characteristic velocity of
the system) in the following sense: the
commutator of $A(t)$ with any operator $B$ is exponentially small
if $B$ is at least distance $t/c_1$ from $A(0)$.  This applies
in particular for $A=u_s$.  Thus, in the definition
(\ref{weq}) of $U$, the operators $u_{s'}^+(i\tau)$ are local in the same
sense: Eq.~(\ref{tdef}) gives $\tilde u_{s'}^+(i\tau)$ as an integral over
$t$ of $u_{s'}(t)$ and for $t>>t_q$ the integral is cut off exponentially,
while for $t\sim t_q$ the $u_{s'}(t)$ are local up to length of order
$t_q/c_1$.  Then, we define $O(s)$ by the unitary transformation
$\t V(s)$.  We view this unitary transformation as defining a fictitious
time evolution with time parameter $s$ and Hamiltonian given by
the exponent of Eq.~(\ref{weq}).  We have just established that this
exponent is local and we can then apply the finite group velocity
result to this evolution to show that $O(s)$ is also local.  The rest
of this subsection consists of a few precise error bounds following
these statements.

We compute the
commutator $[\tilde u^{i+}_s(i\tau),O_j]$ where $O_j$ is some operator
which acts only on site $j$, and
where $\tilde u^{i+}_s(i\tau)$ is defined following Eq.~(\ref{tdef})
taking $A=u^i_s$.
We separate the integral over times $t$ into
times with $t<c_1 l$, where $l=d(i,j)-R$, and times with
$|t|>c_1 l$.  For the first set of times, we have
$(2\pi)^{-1} \int_{|t|\leq c_1 l} {\rm d}t (it+\tau) ||[u_s(t),O_j]||\leq
(2\pi)^{-1} ||u^i_s|| ||O|| \exp(-l/\xi_C)$.  For the
second set of times, we have
$(2\pi)^{-1} \int {\rm d}t (it+\tau) ||[u_s(t),O_j]||\leq
\pi^{-1} ||u^i_s|| ||O|| (\sqrt{2 \pi }t_q/ c_1 l)
\exp[-(c_1 l/t_q)^2/2]$.  Thus, for large $l$,
we find $||[\tilde u^{i+}_s(i\tau),O_j]||$ is exponentially decaying in $d(i,j)$
with decay length $\xi_C$.  Note that 
$\exp(-l/\xi_C)\approx \exp[-(c_1 l /t_q)^2/2]$ for
$l\approx 2 (t_q/c_1)^{2}/\xi_C$.

Now, here is the trick.  We regard Eq.~(\ref{weq}) as defining the
``time" evolution of states, where the parameter $s'$ is an effective
time parameter and 
$D=i\int_0^{\infty} {\rm d}\tau 
\exp[-(\tau/t_q)^2/2]
[\tilde d_{s'}^+(i\tau)-
h.c.]$ is some effective $s'$-dependent ``Hamiltonian", so
that $\t V(s)={\cal S}'\exp[-i\int_{0}^s {\rm d}s' D]$.
We write
$D=\sum_i D^i$, where $D^i=\int_0^{\infty} {\rm d}\tau
\exp[-(\tau/t_q)^2/2]
[\tilde u_{s'}^{i+}(i\tau)-
h.c.]$.
Then, $||[D^i,O_j]||\leq ||O_j|| F(d(i,j))$, where
the function $F(l)$ is 
exponentially decaying as $\exp[-l/\xi_C]$ for
large $l$ and decaying as $(\sqrt{2\pi}t_q/c_1 l)\exp[-(c_1 l/ t_q)^2/2]$
for small $l$.

This exponential decay is in fact good enough to prove the finite group
velocity result\cite{LR7251} using $D$ as an effective
Hamiltonian; it is not necessary that $D^i$ act only
on sites within some finite range, but an exponential decay also suffices.
Following\cite{H0431}, define $G_i$ by the differential equations, for $s>0$,
$\partial_s G_i(s)=\sum_j G_j(s) F(d(i,j))$ with initial conditions 
$G_i(0)=2$ if $i \in S_O$ and $G_i(0)=0$ otherwise.  Then, one can show that
$||[O(s),O_j]||\leq ||O|| ||O_j|| G_j(s)$.  Solving the equations for $G_j$, one
arrives at the bound (\ref{le}).

One can define the operator $O_{trunc}$ by 
setting $\t V_{\rm trunc}(s)={\cal S}'\exp[-i\int_{0}^s {\rm d}s'
{\rm Tr}_{j,d(O,j)\geq l}(D)]$, where the trace is the trace of operator 
$D$
over sites $j$ with $d(O,j)\geq l$.  Then, set $O_{trunc}(s)=\t V_{trunc}(s)
O \t V_{trunc}(s)^{\dagger}$, getting Eq.~(\ref{otb}).

\subsection{Approximation Result}
The idea behind the proof of the approximation result is that,
for any site $i$, the difference between $\tilde u_s^{i+}(i\tau)$ and
$u_s^{i+}(i\tau)$ can be made small by taking large enough $t_q$.
Here, $u_s^{i+}$ is the positive energy part of $u_s$ and $\tilde u_s^{i+}$
is defined following Eq.~(\ref{tdef}) with $A=u_s^i.$  The difference
between these two is closely related to $e^i$, as given below.
Eq.~(\ref{weq}) involves summing over all sites $i$, but sites $i$ which are
sufficiently far from $O$ will turn out to have little effect on defining
$O(s)$.  Thus, the task in this subsection is to figure out how much difference
there is between 
$\tilde u_s^{i+}(i\tau)$ and $u_s^{i+}(i\tau)$, and then sum that error over
sites near enough to $O$, giving the difference between the quasi-adiabatic
continuation and the adiabatic continuation.  Since the adiabatic continuation
preserves expectation values, this will give an estimate in the error in
the expectation values.  We now do this carefully.

The proof of Eq.~(\ref{ce})
involves defining an additional operator and using triangle inequalities.
We define
\begin{eqnarray}
\label{lqo}
\t V_l(s)
={\cal S}'\exp\{-\int_0^{s}
{\rm d}{s'}
\int_0^{\infty} {\rm d}\tau 
\sum_i \exp[-(\tau/t_{q_i})^2/2]
\times \\ \nonumber
[\tilde u_{s'}^{i+}(i\tau)-
h.c.]
\},
\end{eqnarray}
where now $t_{q_i}$ may depend on $i$ and we define $\tilde u_{s}^{i+}(i\tau)$ by
$\tilde u_s^{i}(t)=u_s^i(t) \exp[-(t/t_{q_i})^2/2]$, again using the
$t_{q_i}$ which depend on $i$.

We then pick $t_{q_i}=t_q$ for $d(i,O)\leq 2 l$, and
$t_{q_i}=t_q+c_1[d(i,O)-2 l]$ otherwise.  Define
$O_l(s)=\t V_l(s) O \t V_l(s)^{\dagger}$.
Thus, $\t V_l(s)$ has a $t_{q_i}$ which increases the further one
gets from operator $O$.  While the operator $\t V(s)$ would create
local excitations everywhere acting on a ground state, $\t V_l(s)$ only
creates local excitations near $O$.

Using a triangle inequality,
$|\langle \Psi_{\rm low,s} |Q_0(s) O(s) Q_0(s)^{\dagger}-O_{\rm adiab}(s)|
\Psi_{\rm low,s} \rangle|
\leq
|\langle \Psi_{\rm low,s} |Q_0(s) O(s) Q_0(s)^{\dagger}-
Q_0(s) O_l(s) Q_0(s)^{\dagger} |\Psi_{\rm low,s} \rangle|
+|\langle \Psi_{\rm low,s} |
Q_0(s) O_l(s) Q_0(s)^{\dagger} -O_{\rm adiab}(s)| \Psi_{\rm low,s} \rangle|$.
The difference between $\t V_l(s)$ and $\t V(s)$ is the excitations
far from $O$.  That is, if for small $s$ $\t V(s) V(s)^{\dagger}=
1+s \sum_i e^i+...$ and $\t V_l(s) V(s)^{\dagger}=1+s \sum_i e^i_l+...$, then $e^i_l$ and
$e^i$ differ only for $i$ far from $O$.
However, for $i$ far from $O$, the $e^i$ 
commute through $O$ (as discussed physically
before) and so it is possible to bound the difference
$|\langle \Psi_{\rm low,s} |Q_0(s) O(s) Q_0(s)^{\dagger}-
Q_0(s) O_l(s) Q_0(s)^{\dagger} |\Psi_{\rm low,s} \rangle|$.
Precisely,
to bound the first difference, we note that the difference between the definition
of $O_l(s)$ and $O(s)$ has to do terms $u_s^i$
with sites $i$ which are at least a distance
$2 l$ from $O$.  Using the locality bound, one can bound
the commutator of $O(s)$ and $O_l(s)$ with $\tilde u_s^i$ for these sites.
This gives
$||O(s)-O_l(s)||\leq c_3(s)$, where
\begin{eqnarray}
\label{difb}
c_3(s)=X_3 \sum_{j,d(i,j)\geq 2 l} 
{\rm max}(\exp[-d(j,O)/\xi_C],\\ \nonumber
\exp[-(d(j,O)/l_{q_j})^2/2]),
\end{eqnarray}
for some constant $X_3$ and where $l_{q_j}$ is of order $c_1 t_{q_j}$.

We now bound the difference
$|\langle \Psi_{\rm low,s} |
Q_0(s) O_l(s) Q_0(s)^{\dagger} - O_{\rm adiab}(s)| \Psi_{\rm low,s} \rangle|$.
To do this, it suffices to bound 
$|\langle \Psi_{\rm low,s} Q_0(s) \t V_l(s) -\langle\Psi_{\rm low,s} V(s)|$.
This is equal to
$|\langle \Psi_{\rm low,0} V(s)^{\dagger} Q_0(s) \t V_l(s) - 
\langle \Psi_{\rm low,0}|$.
The operator $\t V_l(s)$ is equal to
${\cal S'} \exp\{-\int_0^s {\rm d}s' 
(\partial_{s'} \t V_l(s')) \t V_l(s')^{\dagger} \}$.
This equals
$V(s) 
{\cal S'} \exp\{-\int_0^s {\rm d}s' 
V(s')^{\dagger} [ (\partial_{s'} \t V_l(s'))
\t V_l(s')^{\dagger} - 
 (\partial_{s'} V(s')) V(s')^{\dagger} ] V(s) \}$.
Thus, 
\begin{eqnarray}
\label{grp}
V(s)^{\dagger} Q_0(s) \t V_l(s) \\ \nonumber
=[V(s)^{\dagger} Q_0(s) V(s)] \times \\ \nonumber
{\cal S'} \exp\Bigl\{-\int_0^s {\rm d}s' 
V(s')^{\dagger} e_l(s')
V(s') \Bigr\},
\end{eqnarray}
where
\be
e_l(s')=(\partial_{s'} \t V_l(s'))
\t V_l(s')^{\dagger} V(s') - 
 (\partial_{s'} 
V(s')) V(s')^{\dagger}.
\ee

 We have grouped the
operators $V(s)^{\dagger} Q_0(s) V(s)$ together in (\ref{grp}) for a reason:
the matrix $Q_0(s)$ is an operator between the low energy states of
$\cH_s$ so
therefore the operator $V(s)^{\dagger} Q_0(s) V(s)$ is an operator
between the low energy states of $\cH_0$.  Now we turn to the exponential
${\cal S'} \exp\{-\int_0^s {\rm d}s' 
V(s')^{\dagger} e_l(s')
 V(s') \}$.  We want to show that
this operator is also equal to, up to some bounded error, an operator between
the low energy states of $\cH_0$.

Define $P_{high}$ to project onto  the high energy states of $\cH_0$.
Then, we can pick $Q_0(s)$ such that
\begin{eqnarray}
\label{ndd}
|\langle \Psi_{\rm low,0} V(s)^{\dagger} Q_0(s) \t V_l(s) - \langle
\Psi_{\rm low,0}| \\ \nonumber \leq
\int_0^s {\rm d}s' |
\langle \Psi_{\rm low,0} V(s')^{\dagger} e^i_l(s')
V(s') P_{high} |,
\end{eqnarray}
We will bound the integral of Eq.~(\ref{ndd}) below; combining this bound
with Eq.~(\ref{difb}) will give Eq.~(\ref{ce}).

Using linear perturbation theory, 
\be
\label{lpt}
 (\partial_{s'} 
V(s')) V(s')^{\dagger}
=-\int_0^{\infty}{\rm d}\tau
[u_s^+(i\tau)-h.c.]+P,
\ee
where
$P$ only has non-vanishing matrix elements between states of the same
energy: $P_{ab}=0$ if $E_a\neq E_b$.

The error $e_l(s)=-P+\sum_i e^i_l(s)$ where
\begin{eqnarray}
e^i_l(s)=
-\int_0^{\infty}{\rm d}\tau
[\exp[-(\tau/t_{q_i})^2/2]\tilde u_s^{i+}(i\tau)-u_s^{i+}(i\tau)-h.c.].
\end{eqnarray}
For $s=0, 
e^i_l(s)=e^i_l$ defined above.  

We now bound the projection into the high energy sector
$|\langle \Psi_{\rm low,s'} e^i_l(s') P_{\rm high} |$.
We can show by performing
some elementary integrations\cite{H0402} that, for any eigenstate $\Psi_{a,s}$
with energy $E_a>\tau/t_{q_i}^2$,
$|\langle \Psi_{\rm low,s}|
\exp[-(\tau/t_{q_i})^2/2]\tilde u_s^{i+}(i\tau)-u_s^{i+}(i\tau)|\Psi_{\rm a,s}\rangle|
\leq \exp[-(\tau/t_{q_i})^2/2]\exp[-(t_{q_i}E_a)^2/2] |(u_s^i)_{0a}|$,
where
$|(u_s^i)_{0a}|$ is the absolute value of the matrix element of
$u_s^i$ between state $\Psi_{\rm low,s}$ and $\Psi_{a,s}$.
Similarly, for $E_a <\tau/t_{q_i}^2$,
$|\langle \Psi_{\rm low,s}
|\exp[-(\tau/t_{q_i})^2/2]\tilde u_s^{i+}(i\tau)-u_s^{i+}(i\tau)|\Psi_{a,s}\rangle|
\leq \exp[-\tau E_a] 
|(u_s^i)_{0a}|$.  Integrating over $\tau$ and summing over states $\Psi_a$
outside the sector of ground states,
using the bound on density of states, we have
\begin{eqnarray}
|\langle \Psi_{\rm low,s} e^i_l(s)
P_{high}
|^2\leq \int {\rm d}E \rho(E) \times
\\ \nonumber
\{(K/E)2\exp[-(t_{q_i}E)^2]
+K t_{q_i}
\exp[-(t_{q_i}E)^2/2]\sqrt{2\pi})\}^2.
\end{eqnarray}

Summing over sites $i$ and using Eq.~(\ref{ndd})
we obtain the bound
$|\langle \Psi_{\rm low,0} V(s)^{\dagger} Q_0(s) \t V_l(s) - \langle
\Psi_{\rm low,0}| \leq
c_2(s)$ where we define
\begin{eqnarray}
\label{teq}
c_2(s)=s \sum_i 
\Bigl( \int {\rm d}E \rho(E) \{
(K/E)2\exp[-(t_{q_i}E)^2]+ \\ \nonumber K t_{q_i}
\exp[-(t_{q_i}E)^2/2]\sqrt{2\pi}\}^2\Bigr)^{1/2}.
\end{eqnarray}
This completes the calculation.


\begin{thebibliography}{29}
\expandafter\ifx\csname natexlab\endcsname\relax\def\natexlab#1{#1}\fi
\expandafter\ifx\csname bibnamefont\endcsname\relax
  \def\bibnamefont#1{#1}\fi
\expandafter\ifx\csname bibfnamefont\endcsname\relax
  \def\bibfnamefont#1{#1}\fi
\expandafter\ifx\csname citenamefont\endcsname\relax
  \def\citenamefont#1{#1}\fi
\expandafter\ifx\csname url\endcsname\relax
  \def\url#1{\texttt{#1}}\fi
\expandafter\ifx\csname urlprefix\endcsname\relax\def\urlprefix{URL }\fi
\providecommand{\bibinfo}[2]{#2}
\providecommand{\eprint}[2][]{\url{#2}}

\bibitem[{\citenamefont{Kalmeyer and Laughlin}(1987)}]{KL8795}
\bibinfo{author}{\bibfnamefont{V.}~\bibnamefont{Kalmeyer}} \bibnamefont{and}
  \bibinfo{author}{\bibfnamefont{R.~B.} \bibnamefont{Laughlin}},
  \bibinfo{journal}{Phys. Rev. Lett.} \textbf{\bibinfo{volume}{59}},
  \bibinfo{pages}{2095} (\bibinfo{year}{1987}).

\bibitem[{\citenamefont{Wen et~al.}(1989)\citenamefont{Wen, Wilczek, and
  Zee}}]{WWZcsp}
\bibinfo{author}{\bibfnamefont{X.-G.} \bibnamefont{Wen}},
  \bibinfo{author}{\bibfnamefont{F.}~\bibnamefont{Wilczek}}, \bibnamefont{and}
  \bibinfo{author}{\bibfnamefont{A.}~\bibnamefont{Zee}},
  \bibinfo{journal}{Phys. Rev. B} \textbf{\bibinfo{volume}{39}},
  \bibinfo{pages}{11413} (\bibinfo{year}{1989}).

\bibitem[{\citenamefont{Read and Sachdev}(1991)}]{RS9173}
\bibinfo{author}{\bibfnamefont{N.}~\bibnamefont{Read}} \bibnamefont{and}
  \bibinfo{author}{\bibfnamefont{S.}~\bibnamefont{Sachdev}},
  \bibinfo{journal}{Phys. Rev. Lett.} \textbf{\bibinfo{volume}{66}},
  \bibinfo{pages}{1773} (\bibinfo{year}{1991}).

\bibitem[{\citenamefont{Wen}(1991)}]{Wsrvb}
\bibinfo{author}{\bibfnamefont{X.-G.} \bibnamefont{Wen}},
  \bibinfo{journal}{Phys. Rev. B} \textbf{\bibinfo{volume}{44}},
  \bibinfo{pages}{2664} (\bibinfo{year}{1991}).

\bibitem[{\citenamefont{Senthil and Fisher}(2000)}]{SF0050}
\bibinfo{author}{\bibfnamefont{T.}~\bibnamefont{Senthil}} \bibnamefont{and}
  \bibinfo{author}{\bibfnamefont{M.~P.~A.} \bibnamefont{Fisher}},
  \bibinfo{journal}{Phys. Rev. B} \textbf{\bibinfo{volume}{62}},
  \bibinfo{pages}{7850} (\bibinfo{year}{2000}).

\bibitem[{\citenamefont{Moessner and Sondhi}(2001)}]{MS0181}
\bibinfo{author}{\bibfnamefont{R.}~\bibnamefont{Moessner}} \bibnamefont{and}
  \bibinfo{author}{\bibfnamefont{S.~L.} \bibnamefont{Sondhi}},
  \bibinfo{journal}{Phys. Rev. Lett.} \textbf{\bibinfo{volume}{86}},
  \bibinfo{pages}{1881} (\bibinfo{year}{2001}).

\bibitem[{\citenamefont{Wen}(2002{\natexlab{a}})}]{Wqoslpub}
\bibinfo{author}{\bibfnamefont{X.-G.} \bibnamefont{Wen}},
  \bibinfo{journal}{Phys. Rev. B} \textbf{\bibinfo{volume}{65}},
  \bibinfo{pages}{165113} (\bibinfo{year}{2002}{\natexlab{a}}).

\bibitem[{\citenamefont{Wen}(2002{\natexlab{b}})}]{Wlight}
\bibinfo{author}{\bibfnamefont{X.-G.} \bibnamefont{Wen}},
  \bibinfo{journal}{Phys. Rev. Lett.} \textbf{\bibinfo{volume}{88}},
  \bibinfo{pages}{11602} (\bibinfo{year}{2002}{\natexlab{b}}).

\bibitem[{\citenamefont{Sachdev and Park}(2002)}]{SP0258}
\bibinfo{author}{\bibfnamefont{S.}~\bibnamefont{Sachdev}} \bibnamefont{and}
  \bibinfo{author}{\bibfnamefont{K.}~\bibnamefont{Park}},
  \bibinfo{journal}{Annals of Physics (N.Y.)} \textbf{\bibinfo{volume}{298}},
  \bibinfo{pages}{58} (\bibinfo{year}{2002}).

\bibitem[{\citenamefont{Balents et~al.}(2002)\citenamefont{Balents, Fisher, and
  Girvin}}]{BFG0212}
\bibinfo{author}{\bibfnamefont{L.}~\bibnamefont{Balents}},
  \bibinfo{author}{\bibfnamefont{M.~P.~A.} \bibnamefont{Fisher}},
  \bibnamefont{and} \bibinfo{author}{\bibfnamefont{S.~M.}
  \bibnamefont{Girvin}}, \bibinfo{journal}{Phys. Rev. B}
  \textbf{\bibinfo{volume}{65}}, \bibinfo{pages}{224412}
  (\bibinfo{year}{2002}).

\bibitem[{\citenamefont{Wen}(2003{\natexlab{a}})}]{Walight}
\bibinfo{author}{\bibfnamefont{X.-G.} \bibnamefont{Wen}},
  \bibinfo{journal}{Phys. Rev. B} \textbf{\bibinfo{volume}{68}},
  \bibinfo{pages}{115413} (\bibinfo{year}{2003}{\natexlab{a}}).

\bibitem[{\citenamefont{Ardonne et~al.}(2004)\citenamefont{Ardonne, Fendley,
  and Fradkin}}]{AFF0493}
\bibinfo{author}{\bibfnamefont{E.}~\bibnamefont{Ardonne}},
  \bibinfo{author}{\bibfnamefont{P.}~\bibnamefont{Fendley}}, \bibnamefont{and}
  \bibinfo{author}{\bibfnamefont{E.}~\bibnamefont{Fradkin}},
  \bibinfo{journal}{Annals Phys.} \textbf{\bibinfo{volume}{310}},
  \bibinfo{pages}{493} (\bibinfo{year}{2004}).

\bibitem[{\citenamefont{Levin and Wen}(2004)}]{LWstrnet}
\bibinfo{author}{\bibfnamefont{M.}~\bibnamefont{Levin}} \bibnamefont{and}
  \bibinfo{author}{\bibfnamefont{X.-G.} \bibnamefont{Wen}},
  \bibinfo{journal}{cond-mat/0404617}  (\bibinfo{year}{2004}).

\bibitem[{\citenamefont{Motrunich and Senthil}(2002)}]{MS0204}
\bibinfo{author}{\bibfnamefont{O.~I.} \bibnamefont{Motrunich}}
  \bibnamefont{and} \bibinfo{author}{\bibfnamefont{T.}~\bibnamefont{Senthil}},
  \bibinfo{journal}{Phys. Rev. Lett.} \textbf{\bibinfo{volume}{89}},
  \bibinfo{pages}{277004} (\bibinfo{year}{2002}).

\bibitem[{\citenamefont{Moessner and Sondhi}(2003)}]{MS0312}
\bibinfo{author}{\bibfnamefont{R.}~\bibnamefont{Moessner}} \bibnamefont{and}
  \bibinfo{author}{\bibfnamefont{S.~L.} \bibnamefont{Sondhi}},
  \bibinfo{journal}{Phys. Rev. B} \textbf{\bibinfo{volume}{68}},
  \bibinfo{pages}{184512} (\bibinfo{year}{2003}).

\bibitem[{\citenamefont{Hermele et~al.}(2004)\citenamefont{Hermele, Fisher, and
  Balents}}]{HFB0404}
\bibinfo{author}{\bibfnamefont{M.}~\bibnamefont{Hermele}},
  \bibinfo{author}{\bibfnamefont{M.~P.~A.} \bibnamefont{Fisher}},
  \bibnamefont{and} \bibinfo{author}{\bibfnamefont{L.}~\bibnamefont{Balents}},
  \bibinfo{journal}{Phys. Rev. B} \textbf{\bibinfo{volume}{69}},
  \bibinfo{pages}{064404} (\bibinfo{year}{2004}).

\bibitem[{\citenamefont{Wen}(1990)}]{Wrig}
\bibinfo{author}{\bibfnamefont{X.-G.} \bibnamefont{Wen}},
  \bibinfo{journal}{Int. J. Mod. Phys. B} \textbf{\bibinfo{volume}{4}},
  \bibinfo{pages}{239} (\bibinfo{year}{1990}).

\bibitem[{\citenamefont{Wen and Niu}(1990)}]{WNtop}
\bibinfo{author}{\bibfnamefont{X.-G.} \bibnamefont{Wen}} \bibnamefont{and}
  \bibinfo{author}{\bibfnamefont{Q.}~\bibnamefont{Niu}},
  \bibinfo{journal}{Phys. Rev. B} \textbf{\bibinfo{volume}{41}},
  \bibinfo{pages}{9377} (\bibinfo{year}{1990}).

\bibitem[{\citenamefont{Wen}(2004)}]{Wen04}
\bibinfo{author}{\bibfnamefont{X.-G.} \bibnamefont{Wen}},
  \emph{\bibinfo{title}{Quantum Field Theory of Many-Body Systems -- From the
  Origin of Sound to an Origin of Light and Electrons}}
  (\bibinfo{publisher}{Oxford Univ. Press}, \bibinfo{address}{Oxford},
  \bibinfo{year}{2004}).

\bibitem[{\citenamefont{Wegner}(1971)}]{W7159}
\bibinfo{author}{\bibfnamefont{F.}~\bibnamefont{Wegner}}, \bibinfo{journal}{J.
  Math. Phys.} \textbf{\bibinfo{volume}{12}}, \bibinfo{pages}{2259}
  (\bibinfo{year}{1971}).

\bibitem[{\citenamefont{Kogut}(1979)}]{K7959}
\bibinfo{author}{\bibfnamefont{J.~B.} \bibnamefont{Kogut}},
  \bibinfo{journal}{Rev. Mod. Phys.} \textbf{\bibinfo{volume}{51}},
  \bibinfo{pages}{659} (\bibinfo{year}{1979}).

\bibitem[{\citenamefont{Kitaev}(2003)}]{K032}
\bibinfo{author}{\bibfnamefont{A.~Y.} \bibnamefont{Kitaev}},
  \bibinfo{journal}{Ann. Phys. (N.Y.)} \textbf{\bibinfo{volume}{303}},
  \bibinfo{pages}{2} (\bibinfo{year}{2003}).

\bibitem[{\citenamefont{Wen}(2003{\natexlab{b}})}]{Wqoexct}
\bibinfo{author}{\bibfnamefont{X.-G.} \bibnamefont{Wen}},
  \bibinfo{journal}{Phys. Rev. Lett.} \textbf{\bibinfo{volume}{90}},
  \bibinfo{pages}{016803} (\bibinfo{year}{2003}{\natexlab{b}}).

\bibitem[{\citenamefont{Wilson}(1974)}]{W7445}
\bibinfo{author}{\bibfnamefont{K.~G.} \bibnamefont{Wilson}},
  \bibinfo{journal}{Phys. Rev. D} \textbf{\bibinfo{volume}{10}},
  \bibinfo{pages}{2445} (\bibinfo{year}{1974}).

\bibitem[{\citenamefont{Anderson}(1984)}]{pwa}
\bibinfo{author}{\bibfnamefont{P.~W.}\bibnamefont{Anderson}},
  \emph{\bibinfo{title}{Basic Notions of Condensed Matter Physics, Chapter 3}}
  (\bibinfo{publisher}{Addison-Wesley}, \bibinfo{address}{Reading, Mass.},
  \bibinfo{year}{1997}).

\bibitem[{\citenamefont{Lieb and Robinson}(1972)}]{LR7251}
\bibinfo{author}{\bibfnamefont{E.}~\bibnamefont{Lieb}} \bibnamefont{and}
  \bibinfo{author}{\bibfnamefont{D.}~\bibnamefont{Robinson}},
  \bibinfo{journal}{Commun. Math. Phys.} \textbf{\bibinfo{volume}{28}},
  \bibinfo{pages}{251} (\bibinfo{year}{1972}).

\bibitem[{\citenamefont{Hastings}(2004{\natexlab{a}})}]{H0402}
\bibinfo{author}{\bibfnamefont{M.~B.} \bibnamefont{Hastings}},
  \bibinfo{journal}{Phys. Rev. Lett.} \textbf{\bibinfo{volume}{93}},
  \bibinfo{pages}{140402} (\bibinfo{year}{2004}{\natexlab{a}}).

\bibitem[{\citenamefont{Hastings}(2004{\natexlab{b}})}]{H0431}
\bibinfo{author}{\bibfnamefont{M.~B.} \bibnamefont{Hastings}},
  \bibinfo{journal}{Phys. Rev. B} \textbf{\bibinfo{volume}{69}}
  (\bibinfo{year}{2004}{\natexlab{b}}).

\bibitem[{\citenamefont{Affleck et~al.}(1987)\citenamefont{Affleck, Kennedy,
  Lieb, and Tasaki}}]{AKL8799}
\bibinfo{author}{\bibfnamefont{I.}~\bibnamefont{Affleck}},
  \bibinfo{author}{\bibfnamefont{T.}~\bibnamefont{Kennedy}},
  \bibinfo{author}{\bibfnamefont{E.~H.} \bibnamefont{Lieb}}, \bibnamefont{and}
  \bibinfo{author}{\bibfnamefont{H.}~\bibnamefont{Tasaki}},
  \bibinfo{journal}{Phys. Rev. Lett.} \textbf{\bibinfo{volume}{59}},
  \bibinfo{pages}{799} (\bibinfo{year}{1987}).

\bibitem[{\citenamefont{Affleck et~al.}(1988)\citenamefont{Affleck, Kennedy,
  Lieb, and Tasaki}}]{AKL8877}
\bibinfo{author}{\bibfnamefont{I.}~\bibnamefont{Affleck}},
  \bibinfo{author}{\bibfnamefont{T.}~\bibnamefont{Kennedy}},
  \bibinfo{author}{\bibfnamefont{E.~H.} \bibnamefont{Lieb}}, \bibnamefont{and}
  \bibinfo{author}{\bibfnamefont{H.}~\bibnamefont{Tasaki}},
  \bibinfo{journal}{Comm. Math. Phys.} \textbf{\bibinfo{volume}{115}},
  \bibinfo{pages}{477} (\bibinfo{year}{1988}).

\bibitem{haldane} F. D. M. Haldane, Phys. Lett. {\bf 93A}, 464 (1983).


\bibitem{string} M. den Nijs and K. Rommelse, Phys. Rev. B {\bf 40},
4709 (1989); S. M. Girvin and D. P. Arovas, Phys. Scr. T {\bf 27}, 156
(1989); T. Kennedy, J. Phys.: Cond. Matt. {\bf 2}, 5737 (1990).


\end{thebibliography}

\end{document}